%% file: main.tex
\documentclass[]{HLreport}

\usepackage{booktabs}
\usepackage{afterpage}
\usepackage[T1]{fontenc}
\usepackage{textcomp}
\usepackage[numbers,sort&compress]{natbib}
\usepackage{graphicx,mathptmx,amsmath,amsfonts,amscd,amsthm,amssymb,eucal,psfrag,color,subfig,url}

\documentlabel{May 2018}

\begin{document}

\title{R\&D PROPOSAL \\ 
RD51 EXTENSION BEYOND 2018}
\author{\it EDITORS: \\ S. Dalla Torre (INFN Trieste), E. Oliveri (CERN), \\
L. Ropelewski (CERN), M. Titov (CEA Saclay)}
%
\abstract{
\input{short_abstract} 
}

\maketitle
\textbf{List of collaborating Institutes:}
\\
\input{institutes}

\newpage
\tableofcontents
\newpage

\section{Executive Summary}
\input{executive-summary}
\newpage
\section{Introduction}
\input{introduction}
\section{RD51 Legacy, Expertize and Infrastructures}
\label{assets_status}
\input{assets_status}

\section{RD51 extension beyond 2018: the overall scope and objectives}
\input{extension}

\label{extension}
\input{assets_future}

\section{Requests to CERN}
\input{requests}




\end{document}

%% file: short_abstract.tex
The RD51 Collaboration, in charge of the development and dissemination
of MicroPattern Gaseous Detectors (MPGD) since 2008, proposes to extend
its activity, after 2018, for a further five-year term. Since the RD51 initial years, 
the community of MPGD developers and users has grown considerably. 
It is reflected by the many MPGD-based applications 
in high energy and nuclear physics 
experiments as well as in other basic and applied-research fields. 
They rely on the parallel progress of detector concepts and associated technologies.
The cultural, infrastructure and networking support offered by RD51 has  
been essential in this process. The rich portfolio of MPGD 
projects, under constant expansion, is accompanied by novel 
ideas on further developments and applications.
\par
The proposed next term of RD51 activities aims at bringing a number 
of detector  concepts to maturity, initiating new projects and 
continuing the support to the community.
Among leading proposed 
projects are ultrafast, high-rate MPGDs;  discharge-free, high-resolution 
imaging detectors with resistive elements and high-granularity 
integrated electronics; novel noble-liquid detector concepts, 
including electroluminescence in gas bubbles; studies of 
environment-friendly 
counting gases and long-term sealed-mode operation; optical-readout
detectors with radiation-hard imagers for 
fundamental research experiments, radiography and other domains.
\par
The proposed R\&D program is also expected to enrich our basic knowledge 
in detector physics, to form a generation of young detector 
experts - paving the way to new detector concepts and applications. 
The vast R\&D program requires
acquiring additional, up-to-date expertise in advanced technologies.

%% file: institutes.tex
\\ \\
Albuquerque, NM, USA, Department of Physics and Astronomy, University of New Mexico \\
Alessandria, Italy, 	University of Piemonte Orientale and                                                          INFN Torino \\
Amsterdam, Netherlands, NIKHEF\\
Annecy le Vieux, France, LAPP\\
Arlington, TX, USA, University of Texas, Arlington\\
Athens, Greece, INP NCSR "Demokritos"\\
Athens, Greece, National Technical University of Athens\\
Athens, Greece, University of Athens\\
Aveiro, Portugal, University of Aveiro\\
Barcelona, Spain, CNM-IMB (CSIC)\\
Bari, Italy, INFN and University Aldo Moro of Bari\\
Beijing, China, Lab. of Radiation Physics, Tsinghua University\\
Bellaterra, Spain, IFAE Barcelona\\
Bhubaneswar, India, National Institute of Science Education and Research\\
Bogota, Colombia, Centro de Investigaciones, Universidad Antonio Narina\\
Bogota, Colombia, National University of Colombia, Physics Department\\
Bonn, Germany, Rheinische Friedrich-Wilhelms Universit{\"a}t\\
Braunschweig, Germany, Physikalisch-Technische Bundesanstalt\\
Budapest, Hungary, Wigner RCP\\
Bursa, Turkey, Department of Physics, Uludag University\\
Cambridge, MA, USA, MIT\\
Charlottesville, VA, USA, Dept of Physics, University of Virginia\\
Coimbra, Portugal, Laboratorio de Instrumentacao Instrumentacao e Fisica Experimental de Particulas\\
Coimbra, Portugal, University of Coimbra\\
Daejeon, Republic of Korea, Center for Axion and Precision CAPP,                           Institute for Basic Science \\
Darmstadt, Germany, GSI\\
Dubna, Russia, JINR\\
Ferrara, Italy, INFN Ferrara and University of Ferrara\\
Frascati, Italy, INFN Laboratori nazionali di Frascati\\
Freiburg, Germany, Physikalisches Institut     \\                                                                 Albrecht-Ludwigs Universit{\"a}t\\
Garching, Germany, Physik Department E18  \\                                                            Technische  Universit{\"a}t M{\"u}nchen\\
Gatchina, Russia, St Petersburg Nuclear Physics Institute\\
Geneva, Switzerland, CERN\\
Geneva, Switzerland, DPNC, section de physique, Universit\'e de Gen\`eve\\
Gent, Belgium, Dept of Physics and Astronomy, University Gent\\
Gif sur Yvette, France, CEA IRFU (Saclay)\\
Grenoble, France, Institut Max von Laue - Paul Langevin  (ILL)\\
Hamburg, Germany, DESY\\
Hefei, China, State Key Laboratory of Particle Detection and Electronics, University of Science and Technology of China\\
Helsinki, Finland, Helsinki Institute of Physics\\
Kobe, Japan, Kobe University\\
Kolkata, India, Saha Institute of Nuclear Physics\\
Kolkata, India, Variable Energy Cyclotron Centre, VECC\\
Krakow, Poland, AGH University\\
Lanzhou, China, Institute of Nuclear Research, Lanzhou University\\
Lund, Sweden, European Spallation Source\\
Magurele, Romania, IFIN-HH\\
Mainz, Germany, University of Mainz\\
Melbourne, FL, USA, Florida Institute of Technology\\
Mexico City, Mexico, Universidad National Autonoma de Mexico\\
Milano, Italy, University of Milano Biococca\\
Cagliari, Italy, INFN Cagliari and University of Cagliari\\
Moscow, Russia, Institute for Nuclear ResearchResearch, Russian Academy of Sciences\\
Mumbai, India, Nuclear Physics Division, Bhabha Atomic Research Centre\\
M{\"u}nchen, Germany, Max-Planck-Institut f{\"u}r Physik\\
Nantes, France, SUBATECH\\
Napoli, Italy, INFN Napoli  and University Federico II of Napoli\\
Novara, Italy, TERA Foundation\\
Novosibirsk, Russia, Budker Institute of Nuclear Physics and Novosibirsk State University\\
Orsay, France, Laboratoire d'Acclelerateur Lineaire (LAL)\\
Palaiseau, France, Laboratoire Leprince-Ringuet\\
Philadelphia, USA, Temple University\\
Siena, Italy, University of Siena and INFN Pisa\\
Rehovot, Israel, Weizmann Institute of Science\\
Roma, Italy, INFN Roma La Sapienza and University La Sapienza of Roma \\
Roma, Italy, INFN and University Roma 3\\
Roma, Italy, Istituto Superiore di Sanit\`{a}  and INFN Roma La Sapienza\\
Rustrel, France, Low background noise interdisciplinary underground laboratory, LSBB, CNRS\\
Santiago, Spain, Instituto Galego de Fisica de Altas Energias\\
Sao Paolo, Brazil, University of Sao Paolo\\
Seoul, Korea,    Republic of, Korea Detector Laboratory, Korea University\\
Shanghai, China, Jiao Tong University\\
Stony Brook, USA, Physics Department\\
Thessaloniki, Greece, Aristotle University of Thessaloniki\\
Tokyo, Japan, University of Tokyo\\
Tomsk, Russia, Tomsk Polytechnical University\\
Torino, Italy, INFN Torino and University of Torino      \\                                                 
Trieste, Italy, INFN Trieste and University of Trieste\\
Tsukuba, Japan, Tsukuba University of Tsukuba\\
Tucson, AZ, USA, Dept. of Physics, University of Arizona\\
Tunis, Tunisia, Centre National des Sciences et Technologies Nucleaire\\
Upton, NY, USA, Brookhaven National Laboratory\\
Valencia, Spain, Instituto de Fisica Corpuscular (IFIC)\\
Valetta, Malta, University of Malta\\
Valparaiso, Chili, Departamento de Fisica, Universita' Tecnica Federico Santa Maria\\
Wako, Japan, Nishina Center for Accelerator-Based Science, RIKEN\\
Warsaw, Poland, Institute of Plasma Physics and Laser Microfusion\\
Wuhan, China, Huazhong Normal University\\
Zagreb, Croatia, Zagreb University\\
Zaragoza, Spain, University of Zaragoza\\
Zurich, Switzerland, Institute for Particle Physics, ETH Zurich\\ \\

%% file: executive-summary.tex
The RD51 Collaboration, in charge of the development and dissemination
of Micro-Pattern Gaseous Detectors (MPGD) since 2008, proposes to extend
its activity, after 2018, for a further five-year term. This proposal, 
where the Collaboration community expresses firm interest, 
is based on the analysis of the progress in the MPGD sector during 
the RD51 years and the Collaboration contribution to the 
present state-of-the-art; on the legacy in terms 
of expertize and infrastructures that RD51 has built-up over the years; 
on the perspectives for future R\&D projects and the wide dissemination 
of MPGD usage, where the Collaboration can continue offering relevant
cultural and infrastructural support. Therefore, 
the scope and activities for the RD51 extension cluster around 
two main missions: novel R\&D developments towards reaching or overcoming performance limits and the continuation and reinforcement of  the
support actions for the community and MPGD dissemination.
\par
Highlights of the future R\&D scenario based on the ongoing projects  
are central elements of the present proposal. A clear direction 
for future developments is that of the resistive  materials 
and related detector architectures. Their usage improves 
detector stability, making possible higher gain in a single 
multiplication layer, a remarkable advantage for assembly, 
mass production and cost. 
Single layer 
architectures represent also a preferential approach in
hadron sampling calorimetry. 
\par
The frontier of fast and precise 
timing is moved forward by novel developments. Very 
encouraging trends are obtained by coupling gaseous 
detectors and Cherenkov radiators to take advantage 
from the prompt radiation emission. Correspondingly novel electronics and 
front-end circuits are being developed.
\par 
A variety of novel opportunities is offered by MPGD hybridization, a 
strategy aiming to strengthen the detector performance combining the 
advantages offered by a variety of approaches. It is obtained both by 
combining MPGD technologies and by coupling gaseous detectors to different 
detection technologies, as is the case for optical read-out of gaseous 
detectors.
\par 
Contributions to the detector concepts from up-to-date material science are required for several domains: resistive materials, solid-state photon 
and neutron converters, 
innovative nanotechnology components. Material studies can 
contribute to requirements related to low out-gassing, 
radiation hardness, radio 
purity, converter robustness and eco-friendly gases.
\par 
The support actions towards the community are aimed to preserve 
and enrich the present scenario, including: 
networking activity, with focus on training and education, 
further development of dedicated simulation tools, 
advances in electronics tools, continuation of the
collaborative interactions 
with strategic CERN workshops. The need for  new materials 
and technologies
will result in an RD51 action of stimulus to the whole
community of detector developers towards the creation of 
common networking and infrastructures.
\par
The collaboration plans to maintain the current organization, 
characterized by a simple and effective management,
and the existing structure based on the  working groups.
\par 
The Collaboration requests to CERN are moderate, at a level similar 
to the present one, with emphasis to the support of the Gaseous Detector
Development (GDD) laboratory, the collaborative access to 
the EP-DT-EF Micro Pattern Technology (MPT) Workshop and to 
the EP-DT-EF Thin Film and Glass (TFG) Laboratory, as well 
as access to other CERN technical facilities.

%% file: introduction.tex
\subsection{Motivations and mission of the RD51 Collaboration}
The modern photo-lithographic technology on flexible and standard 
PCB substrates has favored the invention, in the last years of the 20th century, 
of novel Micro-Pattern Gas Detectors (MPGD); among them: the
Micro-Strip Gas Chamber (MSGC)~\cite{MSGC}, 
the Gas Electron Multiplier~\cite{GEM} (GEM)
and the MicroMegas (MM)~\cite{MM}. 
Since the very beginning, the goal 
was the development of novel detectors with very high spatial 
($\sim$50~$\mu$m) and time (ns) resolution, large 
dynamic range, high
rate capability (up to $\sim$10$^6$~Hz/$mm^2$), large sensitive area and 
radiation hardness - making them an invaluable tool 
to confront future detector challenges at the frontiers of research.
The dedication of several groups of MPGD developers 
has led to rapid progress, crowned by  new inventions 
and understanding of the underlying operation mechanisms 
of the different detector concepts. MPGDs promised 
to fill a gap between the high-performance but 
expensive solid-state detectors, and cheap but 
rate-limited traditional wire chambers.
Nevertheless,  the integration of MPGDs in large experiments 
was not rapid, in spite of the first large-scale 
application within the COMPASS experiment~\cite{COMPASS} 
at CERN SPS in the early days of the 21st century. 
In COMPASS, telescopes of MMs and GEM trackers demonstrated 
reference performance at particle fluxes of 25~kHz/mm$^2$,
with space resolution better 
than 100~$\mu$m and time resolution around 10~ns. Thus, 
the potentiality of MPGD technologies became evident and the interest 
in their applications has started growing in
the High Energy Physics (HEP) and nuclear physics domains, and beyond. 
Consequently,
it was crucial to consolidate and 
enlarge the dedicated community  to foster
further developments and dissemination of MPGD applications 
in the HEP sector and other fields. 
The RD51 CERN-based technological 
Collaboration~\cite{RD51}, promoted by L. Ropelewski and M. Titov,  
was established to pursue these goals.
\par
\begin{figure}
\begin{center}
\includegraphics[width=1\textwidth]{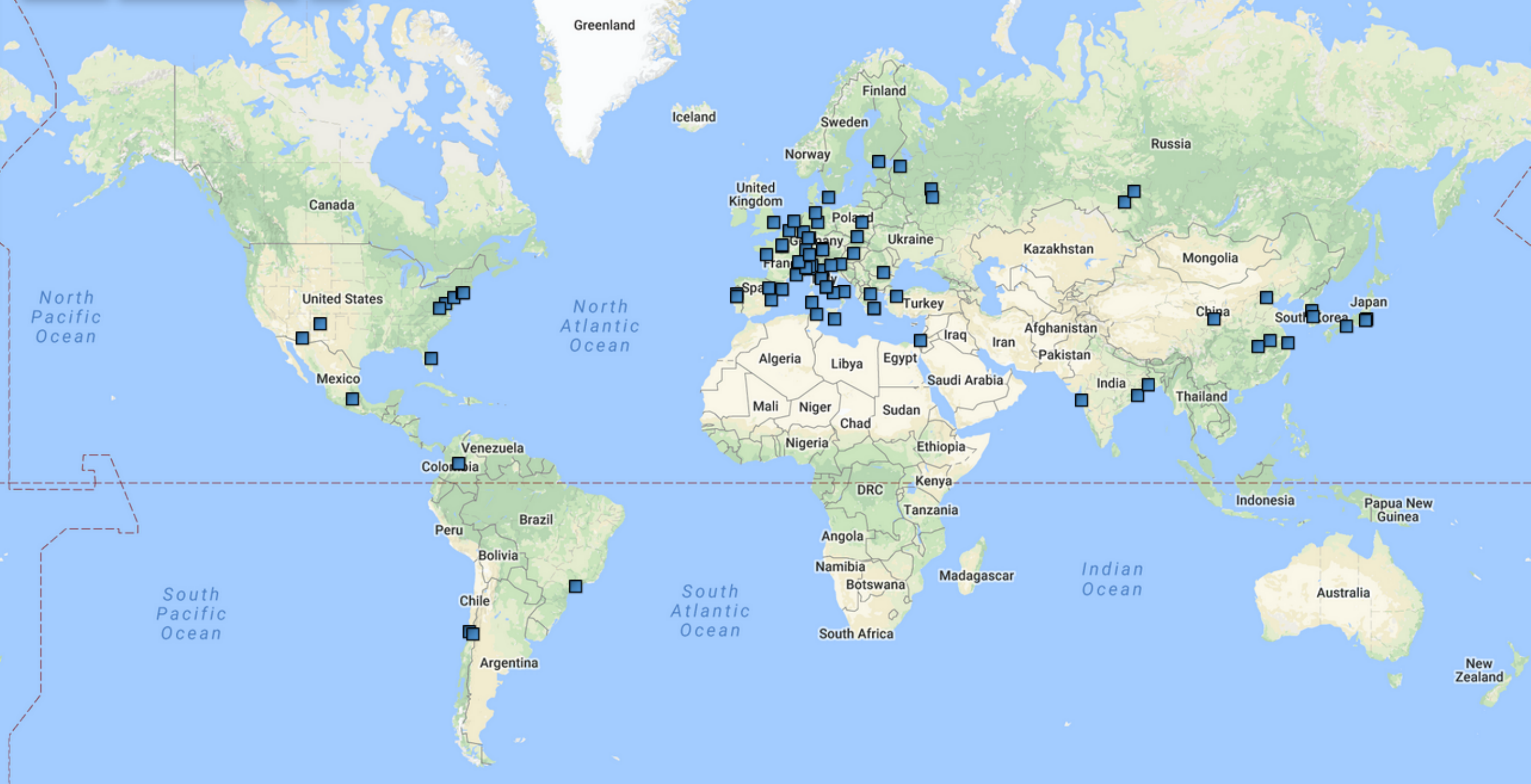}
\end{center}
\caption{\label{fig:RD51Map2018}
\it{Worldwide dissemination of the RD51 collaboration.}}
\end{figure}

RD51 activities started in 2009; the collaboration comprised 
57 Institutions in 21 Countries. Nowadays, the number 
participating Institutions is 
90 in 25 Countries (Fig.~\ref{fig:RD51Map2018}). Initially, the participation was mainly limited to Europe and nearby countries. Over the years, the number of collaborating Institutes from other continents, like from China, India, Japan, USA, has greatly increased - reinforcing the RD51 world-wide vocation and enhancing the geographical diversity and expertise of the MPGD community~\cite{rd51-cern-courier}. 
\\
\par
Since its foundation, the RD51 collaboration has provided 
important stimulus for the development of MPGDs. 
While a number of the MPGD technologies were introduced before 
RD51 was founded, with more techniques becoming available or affordable, 
new detection concepts are still being introduced, 
and existing ones are substantially improved. 
The nature and extent of the Collaboration activities
is reflected in the seven Working Groups (WG), 
transversal to the RD51 activities,  covering 
all relevant R\&D topics: MPGD technologies and novel structures, 
detector characterization, study of 
the physical phenomena and detector simulations, 
dedicated electronics tools for read-out and laboratory studies, 
production and engineering aspects, common test facilities,
and dissemination beyond the HEP 
community - including dedicated education and training.
Since the very beginning, RD51 focused on a broad
networking effort to share and 
disseminate the know-how and the technologies, and the promotion of generic R\&D: 
a seminal activity for the enlargement of the application portfolio. 
Originally created for a five-year term, RD51 was  
prolonged for a second five years term beyond 2013, 
following the LHCC recommendation 
stating: "RD51 is a successful R\&D Collaboration with well-defined 
and important future plans."~\cite{LHCC_June13}.
\subsection{MPGD progress during the RD51 years: the consolidation of
the existing technologies}
Important consolidation of the some better-established MPGD 
technologies has been reached within the RD51 collaboration, 
often driven by the working conditions of large collider experiments.
\par
\begin{figure}
\begin{center}
\includegraphics[width=0.7\textwidth]{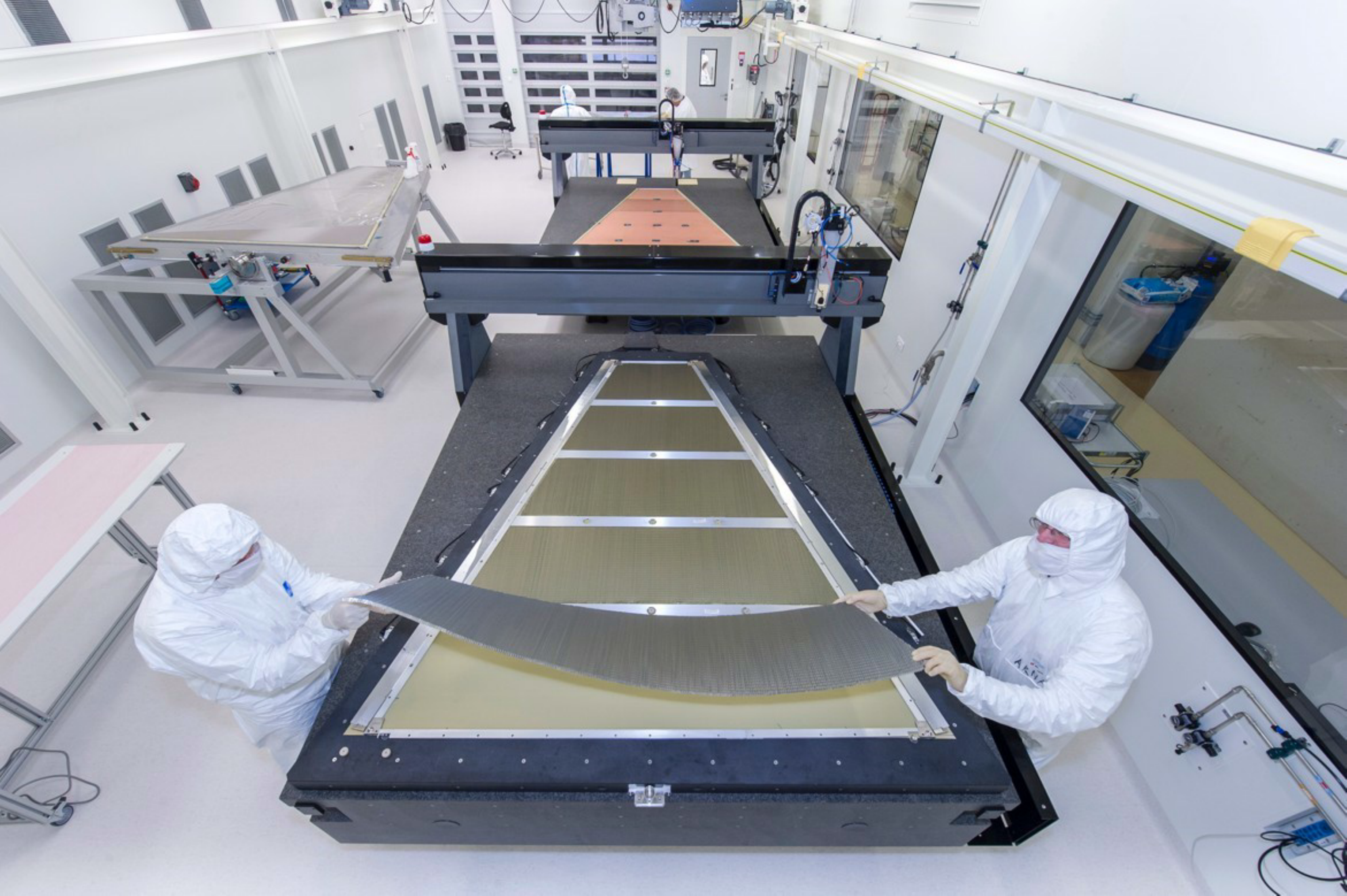}
\end{center}
\caption{\label{fig:ATLAS-NSW-ResistiveMM}
\it{The ATLAS team in the Saclay clean room building a large area resistive micromegas module for the ATLAS New Small Wheel Upgrade~\cite{ATLAS-Picture-NSW-ResistiveMM}.}}
\end{figure}

One of the breakthroughs came with the development 
of MM with resistive anodes for discharge 
mitigation~\cite{resistiveMM}, in the context of the 
ATLAS New Small Wheel (NSW) project~\cite{ATLAS_NSW}. 
Figure~\ref{fig:ATLAS-NSW-ResistiveMM} shows the Saclay 
clean room used for the assembly of the large area final detector. 
This concept 
allows limiting the energy of occasional discharges and
results in a protection of both the detector 
and its front-end electronics and, equally relevant, 
in a substantial dead-time reduction
(time required to re-establish the operational
voltage).
The resistive anodes have been obtained by a variety 
of approaches: photo-lithography, screen-printing technologies
and sputtering. 
\par
\begin{figure}
\begin{center}
\includegraphics[width=1\textwidth]{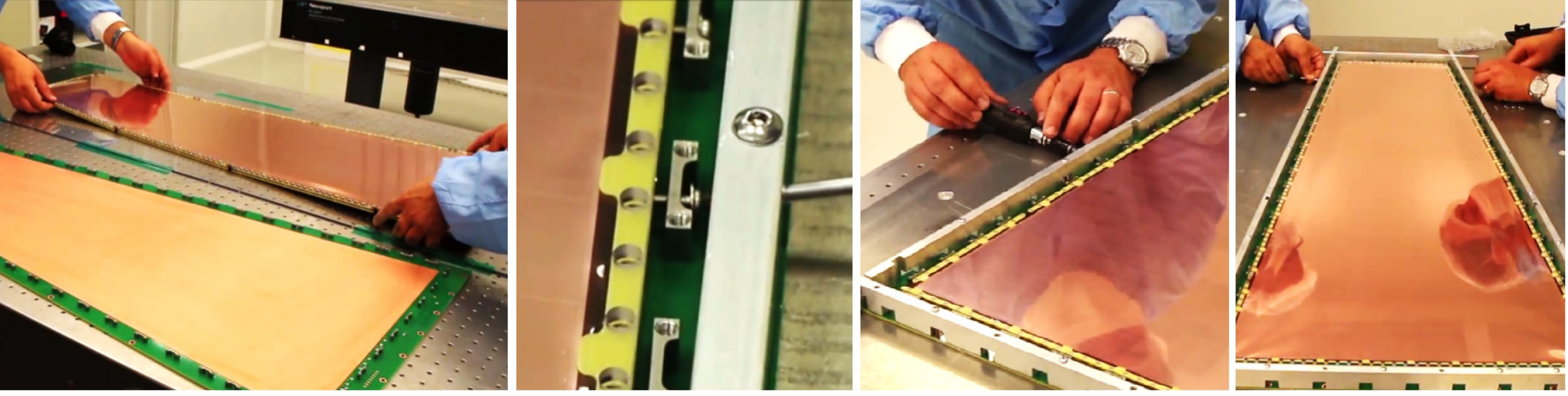}
\end{center}
\caption{\label{fig:CMS-Picture-GE11-GEM}
\it{Left and centre: The NS2 technique, no-stretch no-spacer GEM, developed for the CMS GEM upgrade of the GE1/1 forward muon station ~\cite{CMS-Picture-GE11-GEM}. Right: stretching of the stack of GEM foils during the assembly. Courtesy of the CMS GEM group.}}
\end{figure}
\begin{figure}
\begin{center}
\includegraphics[width=0.6\textwidth]{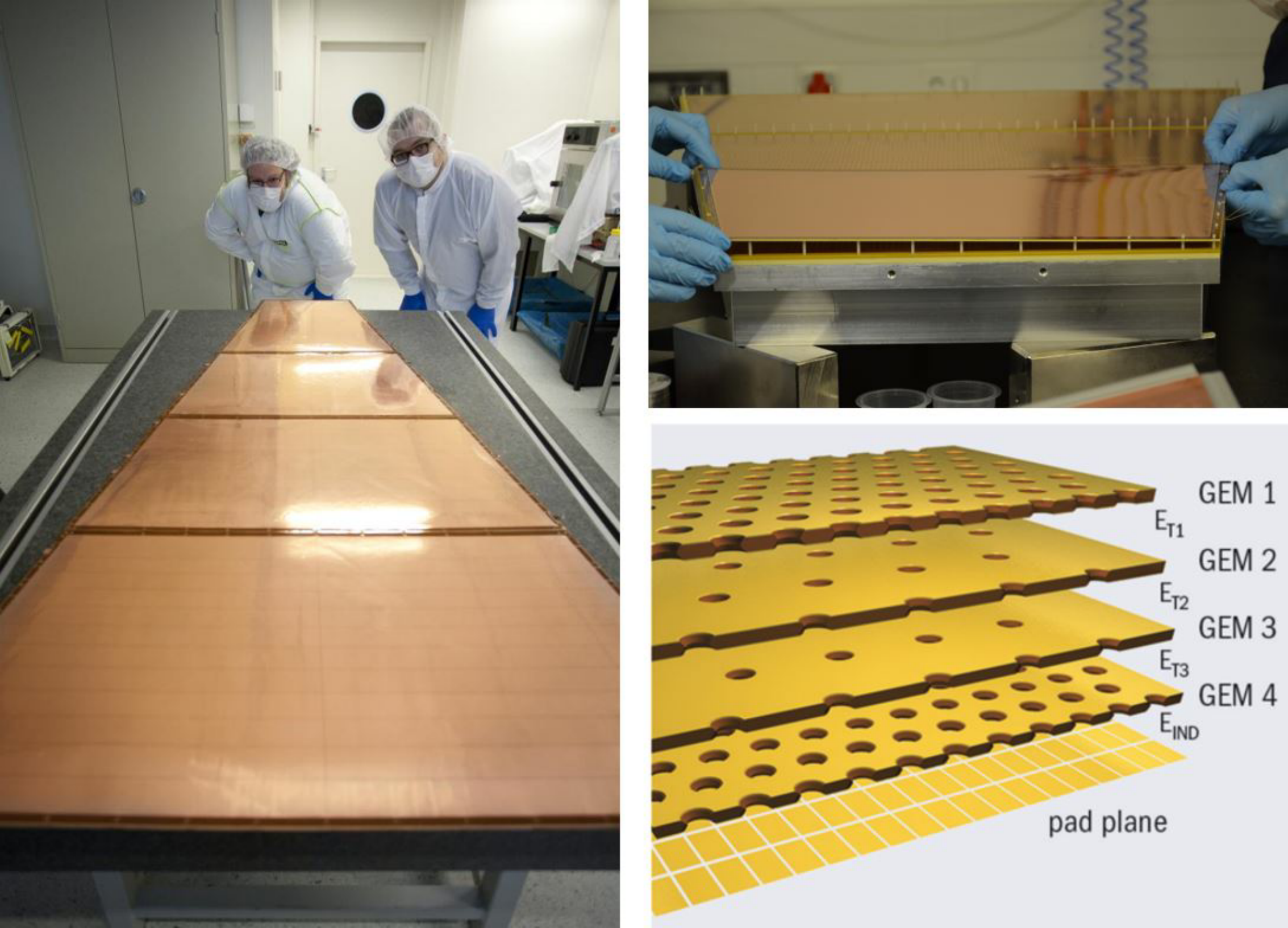}
\end{center}
\caption{\label{fig:ALICE-TPC-GEM}
\it{Left: single mask large-size GEMs of a detector segment for the upgrade of the Time Projection Chamber of the ALICE experiment at CERN~\cite{ALICE-Picture-Left}. Right top: assembly of a framed GEM on top of the pad readout plane. Right Bottom:  Schematic view of the stack of four GEM layer with different pitches optimized for Ions Back Flow suppression~\cite{ALICE-Picture-RightBottom}. Courtesy of the ALICE TPC Upgrade group.}}
\end{figure}

The construction of  GEM-detectors implies 
two main issues: GEM-foil production and 
preservation of the correct spacing between successive GEMs 
in multilayer configurations. Initially GEM foils were produced 
using a double-mask approach with the chemical etching
performed from both foil faces. The difficulty of 
aligning the two masks, limiting the achievable lateral 
size to 50~cm, has led to the development of a single-mask production protocol.
It was initially developed for the upgrade of the 
TOTEM experiment~\cite{single-maskGEM}, further 
used to produce GEM foils for 
the KLOE2 cylindrical GEM-detector~\cite{KLOE-GEM} and that of
the CMS forward muon spectrometer~\cite{CMS-GEM}. The requirement 
to preserve the constant inter-foil spacing in multilayer GEM
detectors of large-size was first fulfilled successfully,
though with small dead-zones,  by adequate spacers; e.g. in 
COMPASS~\cite{COMPASS-GEM} and TOTEM~\cite{TOTEM} trackers. 
Later, the INFN groups 
involved in the construction of the GEM-based trackers for 
LHCb experiment introduced an alternative approach: 
GEM-foil stretching prior to gluing on support frames~\cite{LHCb-GEM}. 
This concept also paved the way to the construction of a 
cylindrical-GEM detector~\cite{KLOE-GEM}. Further extension of 
the stretching technique has been introduced 
for the CMS forward muon spectrometer~\cite{CMS-GEM}: 
to save a relevant fraction of the assembly 
time the foils are mechanically fixed onto the frames, where 
they are mechanically stretched and kept at the 
correct tension without gluing. 
This NS2 technique, no-stretch no-spacer method, presented by R. de Olivera 
at MPGD2013~\cite{GEM-mech-str}, has become the basis of the GEM-detector 
construction for the CMS upgrade. Nowadays, the single-mask GEM technology, 
together with the NS2 technique, simplifies the fabrication 
process, resulting in an important minimization of the production 
time - particularly relevant for large-volume production 
(Fig.~\ref{fig:CMS-Picture-GE11-GEM}). So far, the largest GEM foil 
production is the one ongoing for the upgrade of the ALICE Time Projection Chamber (TPC)~\cite{alice-TPC} 
(Fig.~\ref{fig:ALICE-TPC-GEM}). 
The demanding requirements of a TPC, where fine resolution 
tracking and good dE/dx accuracy are 
equally relevant, has imposed a detailed and 
stringent quality assessment protocol. Therefore, this construction 
effort represents the first fully-engineered large mass-production 
of MPGDs.
\par
\begin{figure}
\begin{center}
\includegraphics[width=0.4\textwidth]{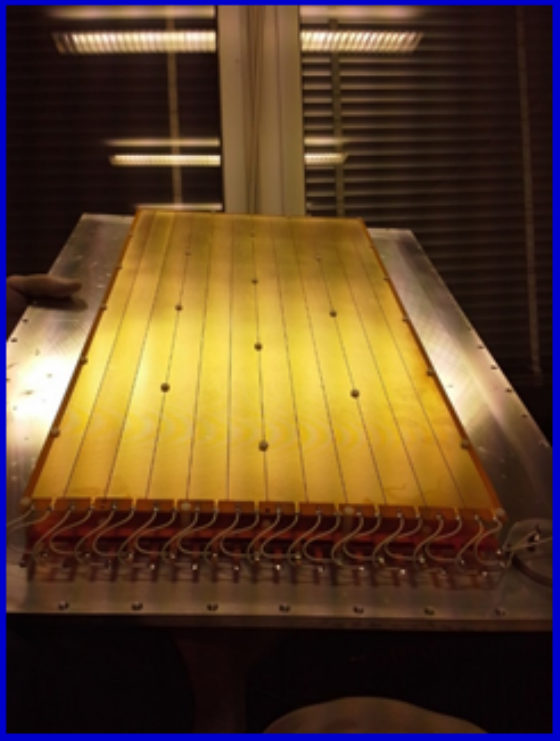}
\end{center}
\caption{\label{fig:compass-thgem}
\it{Picture of a large-size THGEM produced for the gaseous photon detectors of COMPASS RICH. Courtesy of the COMPASS RICH group.}}
\end{figure}

THick GEMs (THGEM), also referred to in the literature as
Large Electron Multipliers  (LEM), have been 
introduced in parallel by several groups at the beginning of 
the 21$^{st}$ century~\cite{THGEM}. They are derived from the GEM design, 
scaling-up $\sim$~10-fold 
the geometrical parameters and changing the production technology. 
The Cu-coated polyimide foil of the GEM multipliers 
is replaced by that of standard 
PCBs (e.g. FR4) and mechanical drilling produces 
the holes (Fig.~\ref{fig:compass-thgem}).
They give access to large avalanche gains and  good rate capabilities. 
THGEM electrodes  can be mass-produced in large sizes 
with standard PCB technology, in spite of the large number of holes: 
some millions per square meter. They 
have intrinsic mechanical stiffness, 
and are robust against damages produced by 
occasional electrical discharges. 
Efforts have resulted in improved electrical stability 
by using a dedicated polishing protocol applied to the 
industrially-produced THGEM electrodes; 
an increased gain uniformity in large-size 
multiplier is reached by careful control of the 
electrode-thickness uniformity~\cite{THGEM-protocol}.
\par 
Advances  have been also achieved on the algorithm side. Exploiting 
the ability of Micromegas and GEM detectors to measure both 
the position and arrival time of the charge deposited in the 
drift gap, a novel method for MM-trackers~\cite{MM-uTPC} 
and later for GEM trackers~\cite{GEM-uTPC} has been developed. 
It allows preserving the space resolution also for inclined particle trajectories: the information collected by thin anode strips is treated as that of a $\mu$TPC. This approach combined  
with a more traditional centroid algorithm permits reaching 
nearly constant resolution of ~$\sim$~100~$\mu$m for trajectory 
inclinations up to 40$^o$.
\subsection{MPGD progress during the RD51 years: novel technologies}
The consolidation of the better-established technologies has been accompanied by the flourishing of novel ones, often specific to well-defined applications. Novel technologies have been derived from MM and GEM concepts, hybrid approaches combining different MPGDs technologies, gaseous with non-gaseous multipliers; others are based entirely on new concepts and architectures.  Novel technologies are illustrated by the several selected examples:
\begin{itemize}
\item 
Coupling the microelectronics industry and advanced PCB technology has been important for the development of gas detectors with increasingly smaller pitch size. An elegant solution is the use of a CMOS pixel ASIC, assembled directly below the GEM or MM amplification structure. Modern "wafer post-processing technology" allows for the integration of a small-scale micromesh grid directly onto a fine-granularity Timepix chip~\cite{timepix}, thus forming integrated read-out of a gaseous detector (InGrid)~\cite{gridpix}. Such a concept allows reconstructing 3D-space points of individual primary electron clusters and serves as an ``electronic bubble chamber''. A breakthrough here is the development of the ILC TPC prototype with a total of 160 InGrid detectors, each 2~$cm^2$, corresponding to 10.5 million pixels~\cite{gridpix-results}, read-out with the RD51 Scalable Readout System (SRS) (Sec.~\ref{Electronics}). New structures, where a GEM foil is facing the Medipix chip~\cite{medipix}, thus forming the GEMpix detector~\cite{GEMpix}, is in use for medical applications as well as for treatment of radioactive wastes~\cite{GEM-rad-hard}. 
\item
GEM geometries with extra electrodes added 
onto one of the two GEM-foil faces in multi-layer configurations 
aim at obtaining a breakthrough in ion-blocking capability - 
a feature required in gaseous photon detectors.
Micro Hole and Strip Plates~\cite{MHSP}
and Cobra~\cite{vis-gas-PMT} architectures have been designed 
and characterized.  
\item
A promising GEM-derived architecture is that of the $\mu$-RWELL~\cite{uR-WELL}, 
where the anode is directly placed at the hole bottom, 
forming the well structure, in order to maximize the collection 
of the avalanche electrons; this single-element structure 
comprises a resistive layer for spark protection.
Several THGEM-based detectors 
have been proposed, including WELL structures, 
with the THGEM electrode being coupled to a readout anode, 
directly or via resistive film or plate for discharge 
damping~\cite{thgem-hcal}.
\item
Initially, GEMs were introduced to act as preamplifier 
in gaseous detectors~\cite{GEM-preamp} and, therefore, the concept of 
the hybrid approach was present since the very 
beginning in MPGD concepts. More recently, 
electron multipliers have been coupled to devices capable of
detecting the luminescent light produced in the amplification process: 
in GEM TPC with optical 
read-out~\cite{GEM-TPC-optRO} and THGEM-based read-out for double-phase 
liquid Ar detectors~\cite{THGEM-Ar-doublephase}.  
A final MM multiplication stage is added to GEM or 
THGEM multipliers in order to control the ion back-flow making use 
of the intrinsic ion trapping capability of the MM principle; 
GEM and MM schemes have been proposed as read-out sensor 
of the ALICE TPC~\cite{alice-TPC}, while THGEMs and MM 
are the basis structure of novel gaseous photon detectors
in COMPASS RICH~\cite{hybrids}. 
\item
The $\mu$PIC~\cite{uPIC} is a fully industrially produced PCB 
including anode strips on one face and orthogonal 
cathode strips on the other one. A regular pattern of 
uncoated zones is present along the cathode strips; an electric 
conductor buried in the thin PCB substrate transfers 
the anode voltage to a "dot"
at the center of each of the uncoated cathode zones: charge 
multiplication occurs
there, under the electric field established between the 
cathode strips and the anode dots. A resistive coating of 
the cathode strips ensures tolerance to occasional discharges.
This technology is easily extensible for the production of 
large areas up to a few square meters.
\end{itemize}
\subsection{MPGD progress during the RD51 years: MPGD applications}
The choice of MPGDs for relevant upgrades of CERN experiments 
indicates the degree of maturity of given detector technologies
for constructing large-size detectors, the level 
of dissemination within the HEP
community and their reliability. 
During the last five years, there have been major MPGDs 
developments for ATLAS, CMS, ALICE and COMPASS upgrades, 
towards establishing technology goals and technical requirements, 
and addressing engineering and integration challenges. 
In parallel, the portfolio of MPGD applications in 
fundamental research has been enlarged in HEP and 
also in other science sectors, as illustrated by the 
following non-exhaustive list of examples.
\par
ATLAS and CMS are building large-area MPGD trackers 
to equip the forward regions, capable of
contributing already at the trigger level.
Large size, about 1$\times$2.5~m$^2$, resistive MM will cover a total surface of 1200~m$^2$ in the ATLAS NSW upgrade~\cite{ATLAS_NSW}, while triple GEM detectors with size up to 1.2$\times$0.45~m$^2$ are being produced for a total GEM-foil surface of about 250~m$^2$ in the CMS GE1/1 upgrade~\cite{CMS-GEM}. The construction of novel four-layer GEM counters designed to upgrade the ALICE TPC  read-out system~\cite{alice-TPC}, for a total GEM-foil surface of about 500~m$^2$, is well advanced.
In 2016, wire-chamber based photon detectors of COMPASS RICH-1 
have been replaced by hybrid MPGDs.
These include
two layers of staggered THGEM and a MM multiplication stage, 
with CsI photocathodes, equipping a total surface of 1.5~m$^2$; 
they accomplish the delicate mission of single photon detection~\cite{hybrids}. 
RD51 has continuously offered to these projects 
consultancy support, 
contributions by dedicated studies on demand;  
e.g. studies of etching processes in GEMs and discharges 
studies in triple-GEM detectors focused to the GEMs for 
the ALICE TPC upgrade. RD51 made also available 
its common test-beam facility at the H4 line at SPS 
and the permanent cosmic ray stand in the Gaseous 
Detector Development (GDD) laboratory, dedicated to 
R\&D studies of the MM for the ATLAS NSW project.
\par
MPGDs are currently being used in a variety of nuclear physics experiments 
and also fulfill the most stringent constraints imposed 
by the future collider facilities, from FAIR and EIC to ILC and FCC.  
At Jlab, 
GEM detectors are deployed in the HallA experiment ~\cite{HallA, HallA2}, 
Super BigBite spectrometer~\cite{Super-BigBite}, 
PRad experiment~\cite{PRad} and in the TPC of 
the BoNuS experiment~\cite{BoNuS}. 
At BNL, the PHENIX-Cherenkov counter has been equipped with CsI-coated 
cascaded-GEM UV detectors in the 
Hadron-Blind Detector (HBD)~\cite{phenix-hbd} and four
GEM trackers in 
the forward direction; moreover GEMs are considered 
for the read-out of the sPHENIX TPC~\cite{sPHENIX}.
\par
MSGC trackers have been 
operated at the DIRAC~\cite{DIRAC} and at HERA-B experiment~\cite{HeraB}. 
A MM detector has been operated at the axion experiment CAST~\cite{CAST}. 
A triple GEM cylindrical detector instruments the vertex region 
of the KLOE2 experiment~\cite{KLOE-GEM} and a second cylindrical detector 
is under construction for BESSIII 
in Beijing~\cite{BESSIII}. Cylindrical MMs have been developed for 
the experiment CLAS12 ~\cite{CLAS12} at JLab and are now one of the 
options for tracking at the future Electron Ion Collider (EIC)~\cite{EIC}. 
A variety of MPGD approaches are being 
considered for TPC read-out at ILC: 
GEMs~\cite{GEM-ILC}, MMs~\cite{MM-ILC} 
and InGrids~\cite{gridPIX-ILC}. 
MM~\cite{mm-hcal}, 
GEM~\cite{gem-hcal} and Resistive plate WELL (RPWELL) 
detectors~\cite{rwell-hcal}
are considered as potential sensing elements in 
Digital Hadron Calorimetry at ILC.
MPGDs are 
developed for low-energy nuclear physics experiments. 
GEMs will read-out a TPC operated at low pressure using light 
gases at the National Superconducting 
Cyclotron Facility (NSCL) in Michigan~\cite{NSCL}, 
MM detectors are the read-out 
sensors of the double-sided TPC of the Neutron Induced 
Fission Fragment Tracking Experiment (NIFFTE) at Los Alamos Neutron 
Science CEnter (LANSCE)~\cite{NIFTE}. 
\par
Working at cryogenic temperatures -~or even within the cryogenic 
liquid itself~- requires optimization to achieve simultaneously high 
gas gain and long-term stability. THGEMs are at the base of 
different developments aiming at novel read-out detectors 
for large-volume noble-liquid TPCs in rare-event experiments. 
Dual-phase liquid Ar TPCs employ LEM/THGEM elements 
in the gas phase as one charge-readout option proposed 
for the DUNE experiment~\cite{DUNE}; they are 
also being investigated in LBNO-DEMO at CERN~\cite{LBNO}.
An alternative, charge read-out concept via avalanche-induced 
electroluminescence in the THGEM holes is considered, 
using Geiger Avalanche Photo Diodes (GAPD), 
in the so-called CRyogenic Avalanche Detector (CRAD)~\cite{CRAD}.
Electroluminescence-photon recording with a gaseous PM (GPM) 
has been recently demonstrated in a dual-phase Xe detector, 
as a potential solution for dark-matter detectors~\cite{arazi}.
THGEMs have been successfully operated 
in a dual-phase Xe detector detecting by PMTs 
the light produced by electroluminescence~\cite{THGEM-Xe}; 
moreover, coating the THGEM with a CsI film, 
it is possible to detect both the ionization 
and the scintillation signal produced in the 
liquid Xe. 
\par
For the field of rare-event searches, Micro-bulk MMs, 
built making use of 
radiopure materials, are studied for the neutrino-less 
double beta decay experiment PANDAX-III~\cite{PANDAX-III} at the 
Jinping Underground Laboratory, China. The NEWAGE0.3b 
Detector at the Kamioka mine, Japan, is developing a 
negative-ion TPC~\cite{NEWAGE0}, using a $\mu$-PIC~\cite{uPIC} detector 
coupled to a pre-amplifying GEM foil. MMs are foreseen for axion search 
in the experiment IAXO~\cite{IAXO}, for WIMPs search at TREX-DM~\cite{TREX-DM}
and for a dark-photon search at P348~\cite{P348}; the experiment 
PADME~\cite{PADME}, also dedicated to a dark photon search, will 
use GEM detectors.
\par
There are number of applications of the MPGDs in the 
neutron detection domain, which include neutron beam 
diagnostics and neutron detection at spallation sources, 
fusion experiments, neutron tomography, and many others.
Neutron imaging by MPGDs 
has been effectively performed by the MSGC-based D20 reflectometer 
at the Institut Laue Langevin (ILL)~\cite{ILL}. 
More recently, neutrons have been detected using 
GEM detectors and polyethylene converter~\cite{ENEA}
at Frascati ENEA Tokamak. High-efficiency n-detectors 
based on GEMs where the neutrons are converted in 
a set of $^{10}$B$_4$C-coated lamellas orthogonal to the 
GEM planes~\cite{BAND-GEM} are under development for applications 
at the European Spallation Source (ESS). 
\par
Beyond fundamental research, MPGDs are in use and 
considered for applications of scientific, social 
and industrial interest; this includes  
the fields of medical imaging, non-destructive tests
and large-size object inspection, homeland security, nuclear plant 
and radioactive-waste monitoring, micro-dosimetry, 
medical-beam monitoring, tokamak diagnostic, 
geological studies by muon radiography and X-ray imaging, 
as illustrated by two recent applications. A telescope 
of MM trackers has been used for the muography of 
the Khufu's pyramid in Egypt, 
contributing to the discovery of a previously not-identified
large void~\cite{pyramid}. A novel method has been proposed 
for measuring  by the GEMPix detector 
the $^{55}$Fe content 
in samples of metallic material activated during operation of 
CERN accelerators and experimental facilities ~\cite{GEM-rad-hard}.
\subsection{MPGD progress during the RD51 years: dissemination}
The RD51 collaboration has been also advancing the MPGD domain 
with scientific, technological and educational initiatives. 
The dissemination of MPGDs beyond fundamental research 
was one of the major new vectors when the continuation 
of the RD51 was approved in 2013. A series of Academy-Industry 
matching events~\cite{AIME}, organized by the RD51 in collaboration 
with HEPTECH~\cite{heptech}, was dedicated to neutron 
detection in 2013~\cite{RD51-n1} and 2015~\cite{RD51-n2}
and to photon detection in 2015~\cite{RD51-photon}. 
These events provided a platform where academic institutions, 
potential users and industry could meet to foster 
collaboration with people interested in MPGD technology. 
Specific R\&D programs are dedicated to applications as, 
for instance, the development of the optical read-out 
of fluorescence light produced by electron multiplication 
in GEM detectors~\cite{GEM-TPC-optRO}, also 
communicated via a CERN Technology Brief~\cite{optical-GEM-TPC}.
More in general, RD51 has contributed to the broad
scientific and technological "know-how" in the MPGD field
by networking 
activity providing financial support to the general-interest 
collaborative efforts and to the specific common facilities and tools. 
Concerning RD51 networking, it is based on the frequent periodical 
meetings of the community: two yearly collaboration meetings, 
inter-spaced by two working mini-weeks. Given the ever-growing interest 
in MPGDs, RD51 re-established an international conference 
series on the 
detectors~\cite{MPGD2009, MPGD2011, MPGD2013, MPGD2015, MPGD2017} 
every second year 
and organizes the schools~\cite{RD51-schools,RD51-schools1,RD51-schools2,RD51-schools3,RD51-schools4} and specialized
workshops~\cite{topical-workshops1,topical-workshops2,topical-workshops3,topical-workshops4}. 
The know-how dissemination is also supported via 
the series of the RD51  Internal Notes~\cite{internal-notes}.
\par
In addition to the support mechanisms and facilities tools, 
further discussed in Sec. 3, the portfolio is rich and diversified, 
including: maintenance and development 
of simulation software dedicated to gaseous 
detectors, development of a complete read-out 
chain designed to operate in a laboratory context, 
also expandable to large read-out systems, 
realization of affordable laboratory instruments 
dedicated to MPGD developments, and more.
Last, but not least, the RD51 community has open access to 
the instrumentation, services and infrastructures 
of the large and well-equipped Gas Detector Development (GDD) 
laboratory at CERN, continuously hosting several 
parallel R\&D activities.
In addition, the common test beam infrastructure at 
the H4 test-beam area at SPS, available usually three 
times a year during the periods of beam availability 
for RD51, allows several groups to investigate in parallel 
their R\&D projects. It is largely maintained by the GDD Team.
\par
\begin{figure}
\begin{center}
\includegraphics[width=0.8\textwidth]{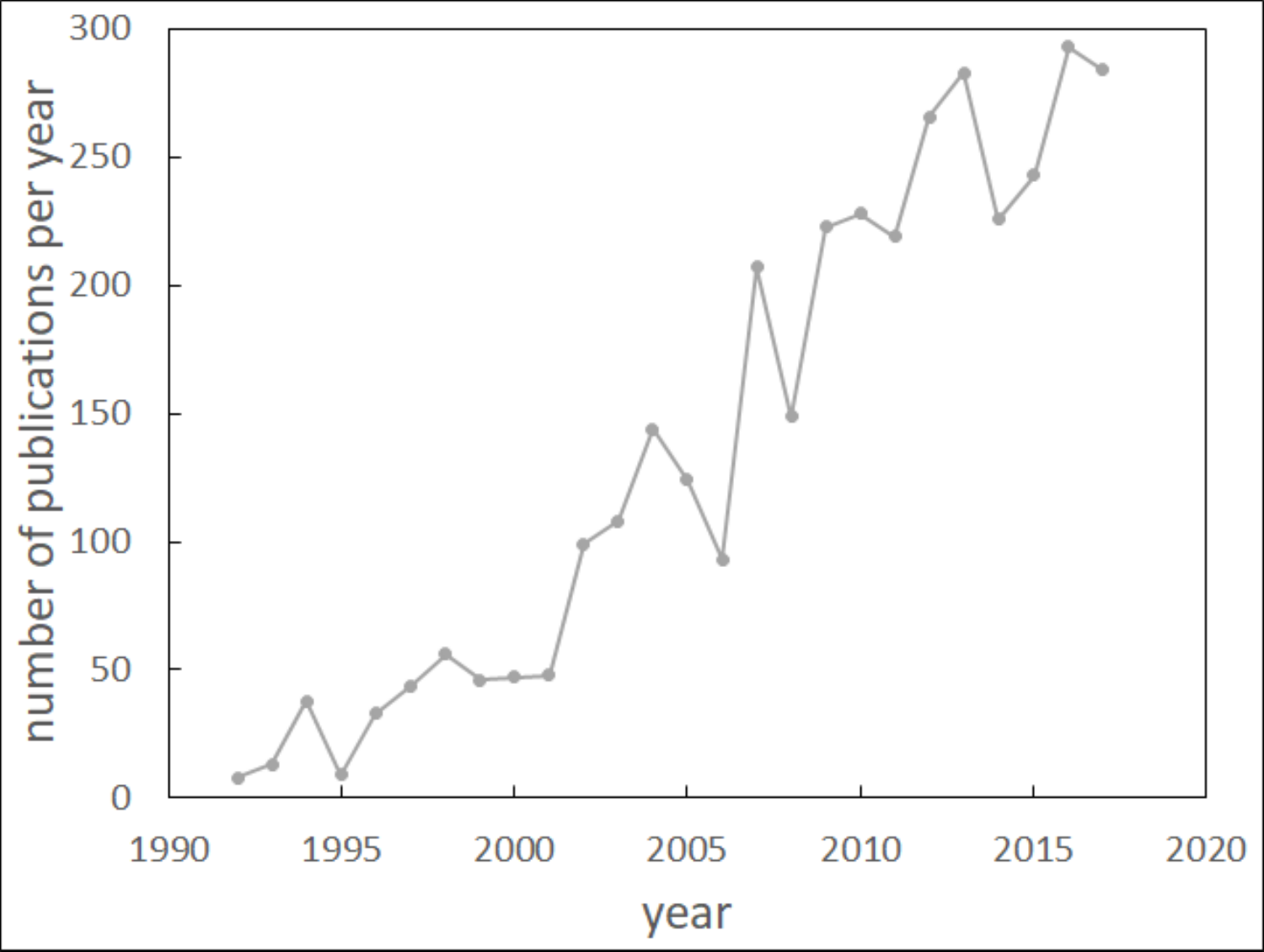}
\end{center}
\caption{\label{fig:publications}
\it{Number of publications dedicated to MPGDs per year; data source: Scopus; estimated error at the 20\% level.}}
\end{figure}

\textit{}
An indicator of the level of interest for MPGDs, 
of the corresponding development effort and the wide 
dissemination of these detectors is the increasing 
number of publications dedicated to MPGDs, 
as illustrated in Fig.~\ref{fig:publications}.

%% file: assets_status.tex
The main objective of the RD51 R\&D program is to advance technological development and application of MPGDs~\cite{RD51-Public}. Since its early stages, the RD51 collaboration has paid attention to building a proper environment for performing high-quality advanced R\&D on MPGDs; it continues to advance the MPGD domain with scientific, technological, and educational initiatives. It is a worldwide open scientific and technological forum on MPGDs, and RD51 has invested resources during ten years in forming expertise, organizing common infrastructures and developing common research tools ~\cite{RD51}. Starting new generic research projects to explore innovative ideas has been strongly boosted by fruitful discussions and exchanges within the broad MPGD community and associated experts. The progress in various R\&D projects has been made possible by open access to facilities and research tools. The RD51 Common Fund has supported the networking activities, projects of common interest. In particular maintenance and development of simulation software, developments in the electronics sector and novel concepts at their starting phase.\\

In several cases, the presence of a trustworthy collaboration behind a given proposal has been also beneficial in proposal submissions to funding agencies. Over the ten RD51 years, many of the achieved results went, in some cases, beyond original expectations. The collaboration is firmly interested in another five-year period extension - to preserve and adapt the existing scenario of successful operation and achievements to future needs. The collaboration is moreover keeping alive, with a programmed and coherent evolution and with a continuous development, several fundamental tools otherwise lost or left to less powerful efficient, isolated individual efforts. This section summarizes the RD51 legacy, expertize and infrastructures available to the community; a few examples representing the current RD51 collaboration patrimony are also discussed.
\subsection{Community and Expertise}\label{Community-Expertise} 
\begin{figure}
\begin{center}
\includegraphics[width=1\textwidth]{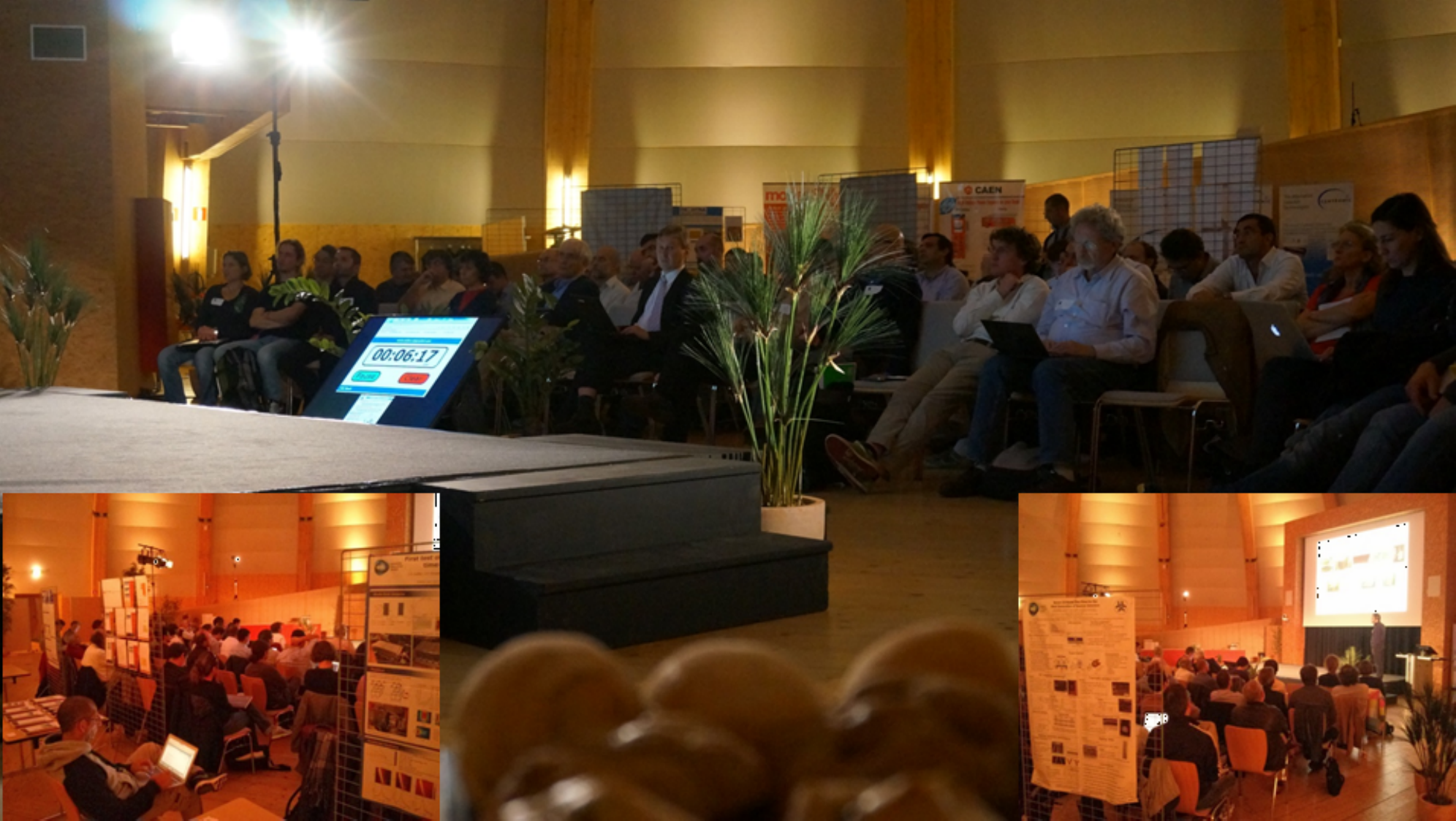}
\end{center}
\caption{\label{fig:RD51AIM}
First  Academy-Industry Matching Event organized by RD51:
\it{Special Workshop on Neutron Detection with MPGDs}~\cite{RD51-n1}. 
}
\end{figure}

The international RD51 community is primarily involved in the development for fundamental-research application and in generic R\&D;  activities 
focused on applications beyond basic research are also continuously growing. The collaboration is widely distributed in terms of institutes and countries. The variety of expertise and the full coverage of all the MPGD concepts and associated techniques is securing a proper balance to address future R\&D challenges, allowing for an open and unbiased exchange of knowledge and information. There is no similar detector R\&D consortium, world-wide,
relying on such freedom and diversity of research groups and their interests.  In this context, every RD51 member is contributing with his (her) own knowledge and acquires, in exchange, all available knowhow and expertise in the community. It has been proven to be an enriching dynamic process for everyone, and especially for the young generation of scientists. Common events (conferences, meetings, workshops, schools, lectures, trainings - Fig.~\ref{fig:RD51AIM}) and common spaces (laboratory and test beam) have been playing an essential role in this fruitful exchange. Envisaging the needs for R\&D activities on MPGD, RD51 has been structured in several working groups. Each of them is focusing on various aspects of gas-detector technologies: detector physics and measurements, new technologies, simulations and modeling, electronics, production techniques and common testing facilities. This structure facilitates interactions within the community and effectively focuses efforts and resources. The Spokespersons, the RD51 Management Board members and the working-group Leaders have the role of linking effectively needs and available expertize and resources; their role is vital for keeping the community together. The aim of the requested extension is to preserve this effective and successful working environment, to support and enrich the existing framework and to form together new important areas of expertise in the context of novel and future detector technologies.
\subsection{Detector physics, simulations and software tools}\label{Simulation}
\begin{figure}
\begin{center}
\includegraphics[width=1\textwidth]{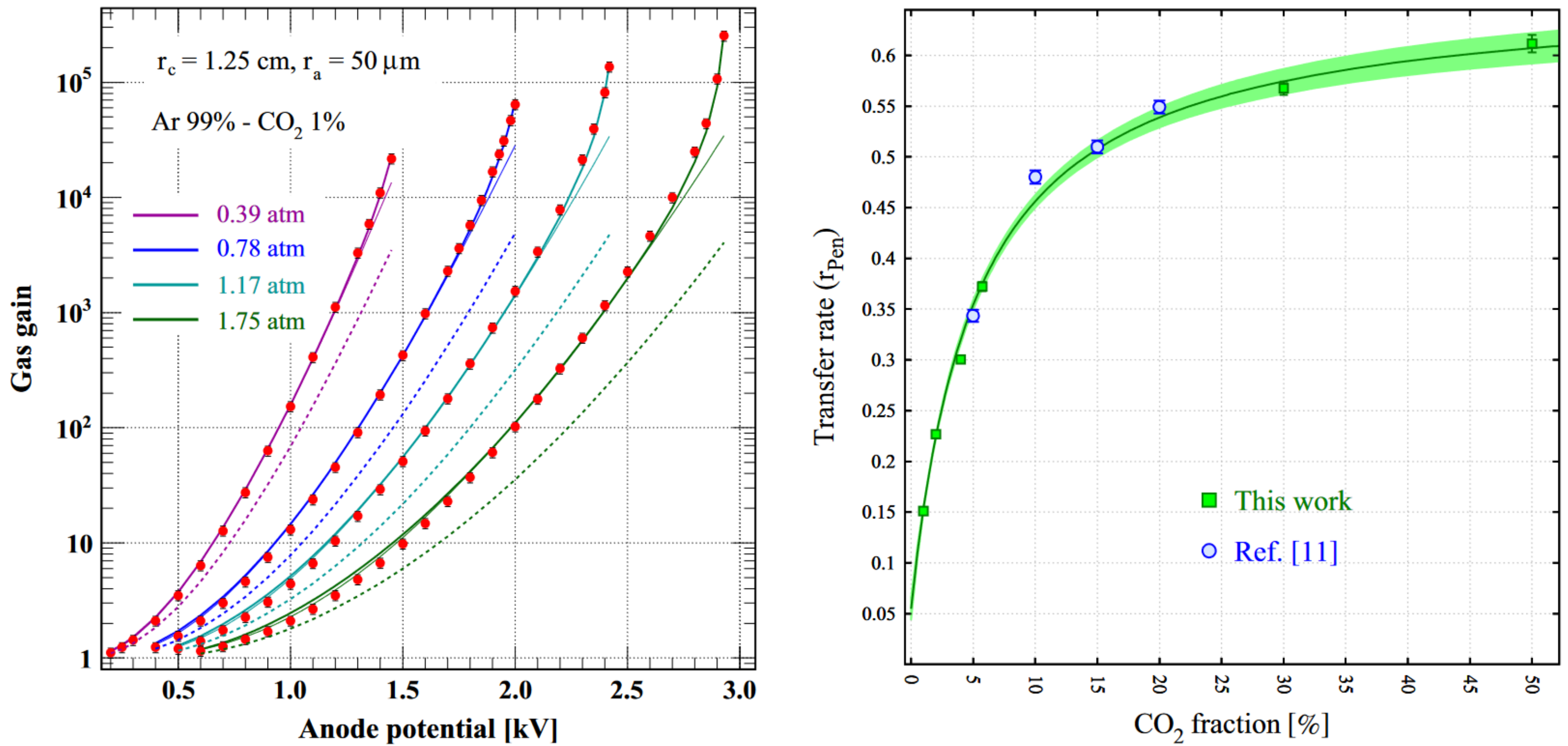}
\end{center}
\caption{\label{fig:Simulation-Penning}
\it{Left plot: Calculated and measured gas gain curves: calculated without corrections (dashed lines), with Penning transfer included (thin straight lines), with Penning transfer and photon feedback included (thick straight lines) and measured data (points). Right plot: Transfer rate in Ar-CO$_2$ as a function of CO$_2$ concentration at mixture
pressure of 1070 hPa~\cite{Simulation-Penning}.}}
\end{figure}

Fast and accurate detector-physics simulations activities have gained  considerable importance, with the increase of the complexity of novel instrumentation. Developed by Rob Veenhof many years ago and continuously updated, Garfield~\cite{Garfield} represents a unique software package for microscopic modeling of detector response. Garfield, together with HEED~\cite{HEED}, Degrad~\cite{Degrad} and Magboltz~\cite{Magboltz}, represents the core of MPGDs simulation tools.  They are all common and open-access to the community. In most of the new MPGD R\&D projects, the suggested concepts are corroborated by these simulation studies; they permit better understanding of the operation mechanisms and expected detector performances. Simulating MPGDs requires an integrated approach of field calculations and charged particle transport, since the field changes substantially over the free path between collisions. Within RD51, an open-source, nearly exact boundary element solver (neBEM) was developed to overcome problems often related to finite-element methods of approximating electrostatic fields in the proximity of electrodes ~\cite{neBEM}.\\ \\ \\ RD51 is supporting and encouraging maintenance and new developments of invaluable simulation tools. It is important to mention that any model simulations require validations by experimental measurements. One example is the dedicated measurement campaign and data analysis program that was undertaken to understand avalanche statistics and determine the Penning transfer-rates in numerous gas mixtures (Fig.~\ref{fig:Simulation-Penning}). Significant efforts were also devoted towards modeling of MPGD performances for particular applications; e.g. studies of electron losses in MM with different mesh specifications for the ATLAS NSW, and GEM electron transparency, charging-up and ion-backflow processes for the ALICE TPC upgrades, and their comparison with experimental data. Individual groups are very often motivated to engage themselves in accurate and, in some cases, very difficult measurements because of the presence of the large RD51 community as the end-user of their efforts.
\subsection{Electronics}\label{Electronics}
\begin{figure}
\begin{center}
\includegraphics[width=1\textwidth]{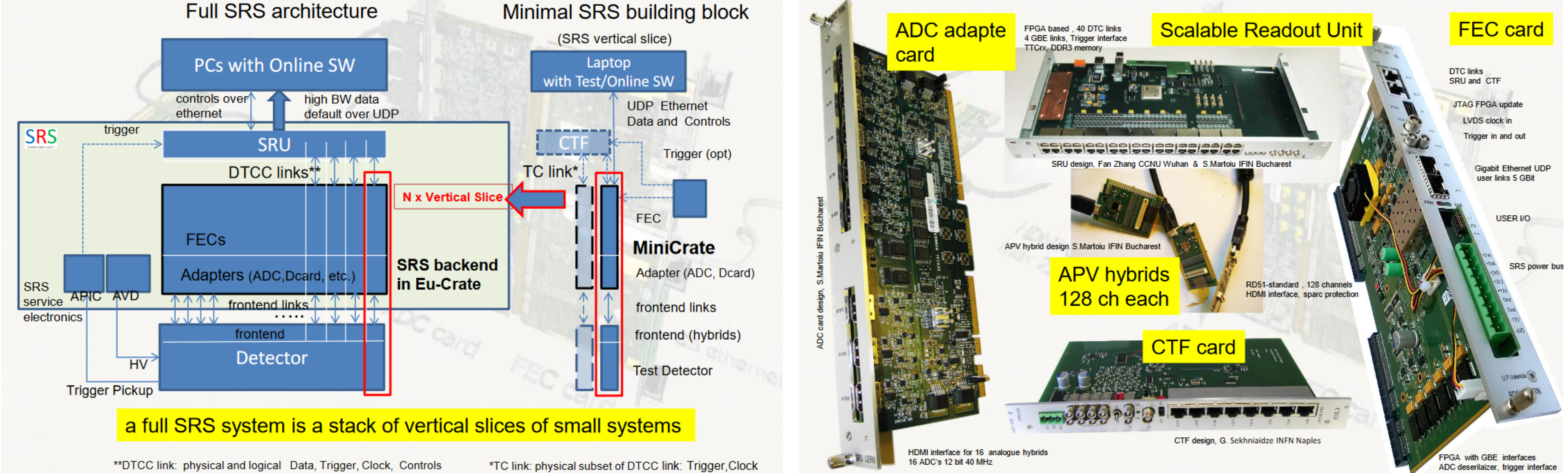}
\end{center}
\caption{\label{fig:SRS}
\it{Left: RD51 Scalable Readout System Architecture. Right: SRS Hardware components.~\cite{SRS}.}}
\end{figure}

    In the front-end electronics and data acquisition systems for MPGDs, detector-electronics integration (e.g. modern gaseous detectors with CMOS pixel readout), and discharge-protection strategies have been among the core missions of RD51.  Developments very specific to our technologies, easier access to otherwise very expensive instrumentation and a large community working with common hardware and software tools are part of motivations behind the RD51 electronics activities. The most prominent RD51 development is a basic multichannel readout system for MPGDs, the so-called Scalable Read-out System (SRS)~\cite{SRS-Paper}(Fig.~\ref{fig:SRS}). It is an "easy-to-use" portable system from detector to data analysis, with read-out software that can be installed on a laptop for small laboratory setups. Its scalability principle allows systems of 100,000 channels and more to be built through the simple addition of more electronic SRS slices, and operated at very high bandwidth using the on-line software of the LHC experiments. The number of SRS systems deployed so far already exceeds 100, with more than 250,000 APV25 front-end channels. It is available through the CERN store or via the CERN Knowledge Transfer office, which also granted SRS reproduction licenses to several companies. Since 2013, SRS has been re-designed in the ATCA industry standard, allowing for much higher channel density and output bandwidth.\\
    
The LHC experiments - ATLAS, TOTEM and CMS - have been using SRS in-house prototypes in laboratory tests in the course of their upgrade projects. The front-end adapter concept of SRS represents another degree of freedom, because any sensor technology typically implemented in multichannel ASICs may be used. Originally designed to operate with APV25, nowadays, the system has been extended allowing implementation of three different ASICs on SRS hybrids as plug-ins for MPGDs: APV25, Timepix and VMM. Following this encouraging experience, SRS has been ported for the readout of photon detectors (ex. SiPMs of NEXT TPC) and tracking detectors (e.g. "InGrid" arrays for ILC TPC). The latter represents a real break-through development of a large-area MPGD with CMOS pixel ASIC allowing reaching the level of integration, compactness and resolving power typical of solid-state pixel devices.\\

The realization of customized and affordable laboratory instruments dedicated to MPGD developments is an additional support R\&D tool. A number of instruments have been developed and are presently at different maturity stages; several of these projects have been supported by the EU AIDA 2020 project~\cite{AIDA2020}. Among them: 
\begin{itemize}
\item active power supply system,  namely Active Voltage Divider (AVD), designed to match the requirements of MPGD architectures including multiple multiplication stages, presently available at prototype level;
\item
femto-ammeter, available in an initial version, while faster readout O(MHz), floating operation and bipolar capability are the planned further improvements; 
\item 
RHIP system~\cite{rhip}, where an array of cheap, fully floating pico-ammeters is read-out via radio connection; 
\item
the APIC unit, a single channel readout unit including pre-amplification, shaper, discriminator, trigger options, presently a mature device ready for commercialization;  
\item
a monitoring unit for environmental  parameters (pressure, temperature, humidity), developed first in an Arduino-based version and, later, in a Raspberry-Pi-based version; 
\item  
a dedicated high voltage system under development  to match MPGD requirements, presently not satisfied by commercial devices; it makes use of commercially available DC-to-DC converters and local intelligence,  its design includes a sophisticated control allowing for a real-time voltage monitoring and current monitoring at the nA level, and performs voltage corrections for temperature and pressure changes. 
\end{itemize}
\subsection{Workshops}
\begin{figure}
\begin{center}
\includegraphics[width=1\textwidth]{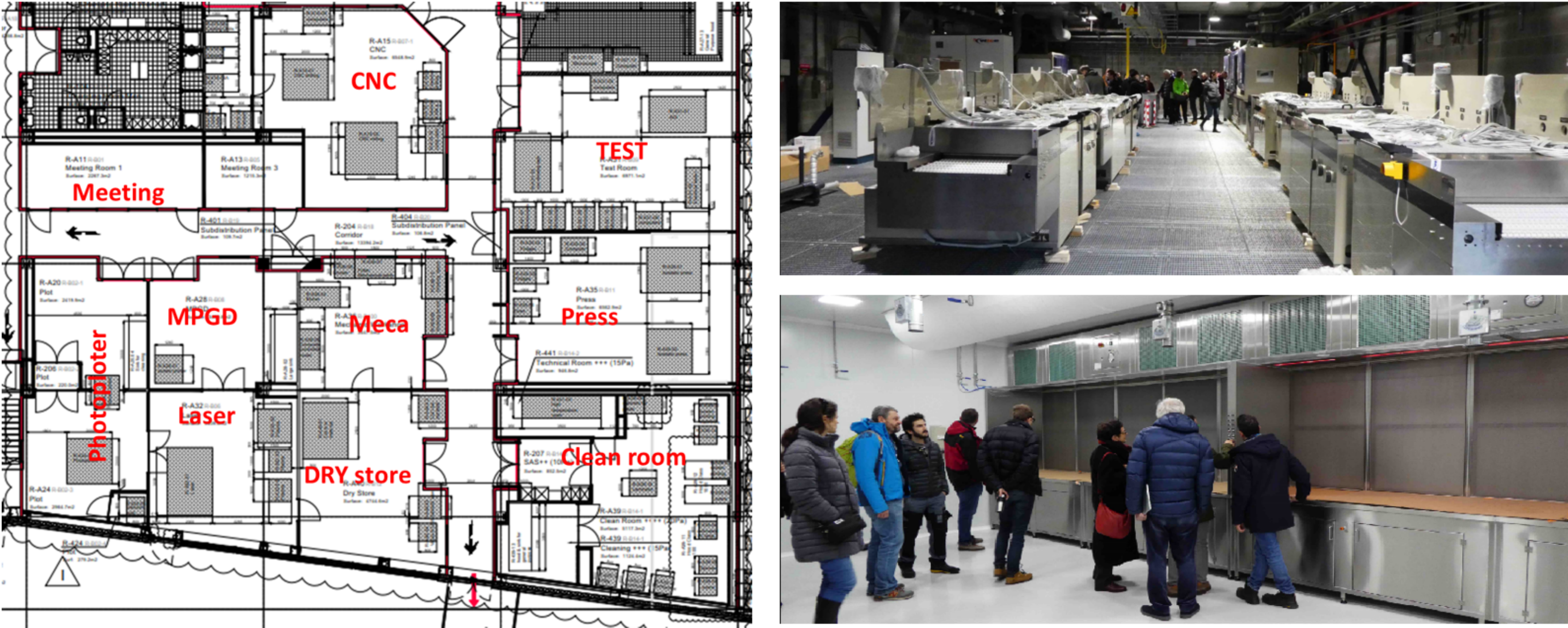}
\end{center}
\caption{\label{fig:MPTWorkshop}
\it{New Micro Pattern Technology (MPT) workshop (building 107) of CERN EP-DT. 
~\cite{MPTWorkshop}}}
\end{figure}
\textbf{MPGD workshops}: For nearly 20 years, the CERN Micro Pattern Technology (MPT) workshop has been a unique facility, where generic R\&D, detector-components production and quality control take place; these allow advancing cost-effective MPGDs manufacturing and their production by industrial processes. Following trends in the large detector units, the RD51 collaboration, representing the needs of the community, proposed an extension of the workshop infrastructure, necessary to manufacture detectors with dimensions up to 2$\times$1~m$^2$, more than doubling the maximum size of MPGDs produced at that time. Approved in 2009 by the CERN management, and promoted within AIDA~\cite{AIDA} and by synergies between CERN and the RD51 community, the installation of new equipment has been completed in 2012 in the old workshop premises. The new, high-tech equipped and up-to-date building will be fully used starting from the fall of 2018 (Fig.~\ref{fig:MPTWorkshop}). 
    This upgrade paved the road to very promising R\&D on the large-area MPGD-based detectors, in a view of large-scale production in industry at a later stage. The current technological status has been attained thanks to the formidable competencies developed over the years, available in the workshop, and to the important support coming from the collaboration and the community. The available expertise is offering the possibility of producing application-tailored detectors and prototypes with a large variety of configurations and complexities, matching a broad range of physics requirements. Scaling up MPGD detectors, while preserving the typical properties of small prototypes, allowed the use of major MPGD technologies in the current upgrades of large experiments, e.g. that of the LHC. More robust solutions, implementing resistive materials and thin films in single-stage detectors, extended the range of possible applications.
   The expertise acquired in the past allowed the workshop to move to full detector production and assembly. With such a widespread dissemination, a technological transfer became essential.  Different approaches have been followed, but in all the cases, the RD51 collaboration has been open to provide support when needed and whenever it had the necessary competencies. Nowadays, the CERN MPT workshop, in collaboration with RD51 and experiments that are using MPGDs on a large scale, serves as a reference point of contact for companies interested in MPGD manufacturing and helps them to reach the required level of competencies. Contacts with some have strengthened to the extent that they have signed license agreements and engaged in a technology-transfer program; more than 10 companies are already producing GEM, MM and THGEMs of reasonable sizes. While, the CERN MPT represents the major infrastructure used by the collaborators for technology advances, several other facilities have also been built. The MPGD workshop at IRFU/CEA Saclay represents another prominent example.\\
\begin{figure}
\begin{center}
\includegraphics[width=1\textwidth]{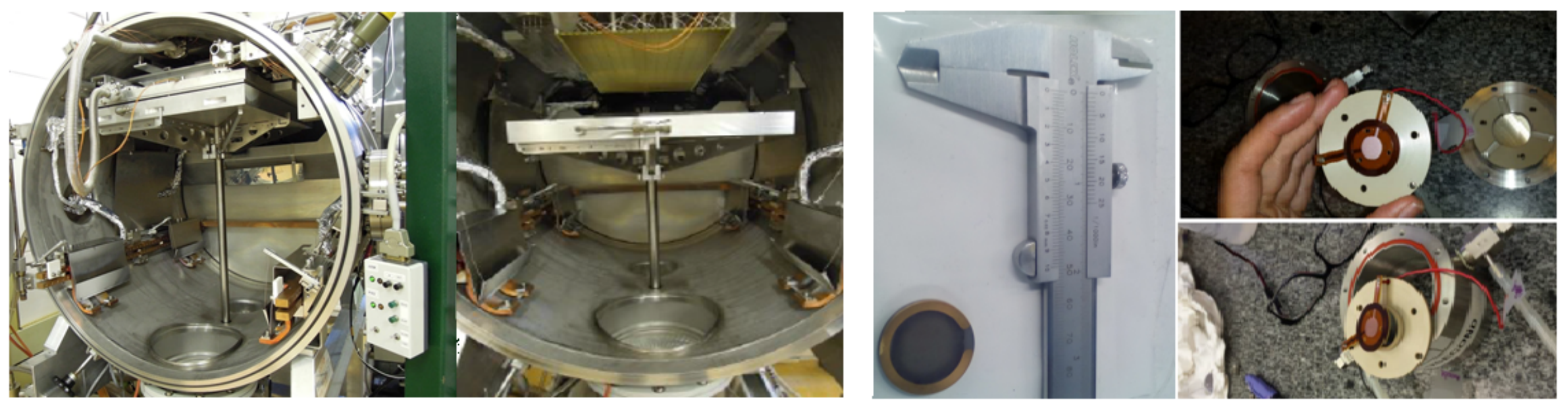}
\end{center}
\caption{\label{fig:TFG}
\it{Left: Large Area CsI Evaporator at the Thin Film and Glass (TFG) EP-DT workshop. The setup has bees used for CsI coating of ALICE HMPID RICH Detector and for the COMPASS THGEM based Upgrade. Right: CsI coating for precise and fast timing R\&D with MPGD (PICOSEC project). 
~\cite{TFG}}}
\end{figure}
\textbf{ Thin-film and Glass laboratory at CERN}:~\cite{TFG-Web} In our community, more specifically in the context of gaseous photodetectors, an important role is played by thin-film coating laboratories capable of depositing photocathodes on detector electrodes. The EP-DT Thin-film and Glass laboratory at CERN (Fig.~\ref{fig:TFG}) is a good example. Large-area CsI photocathodes have been successfully produced for RICH detectors currently deployed at CERN (ALICE HMPID, COMPASS RICH-1 MPGD). A recent R\&D project in RD51 requires CsI deposition for the development of fast and precise timing detectors. Successful results have been achieved in a short time ~\cite{TFG-picosec} thanks to the availability of this facility and its long-standing expertise. 
\subsection{Common space and common test facilities}
\begin{figure}[ht]
\begin{center}
\includegraphics[width=1\textwidth]{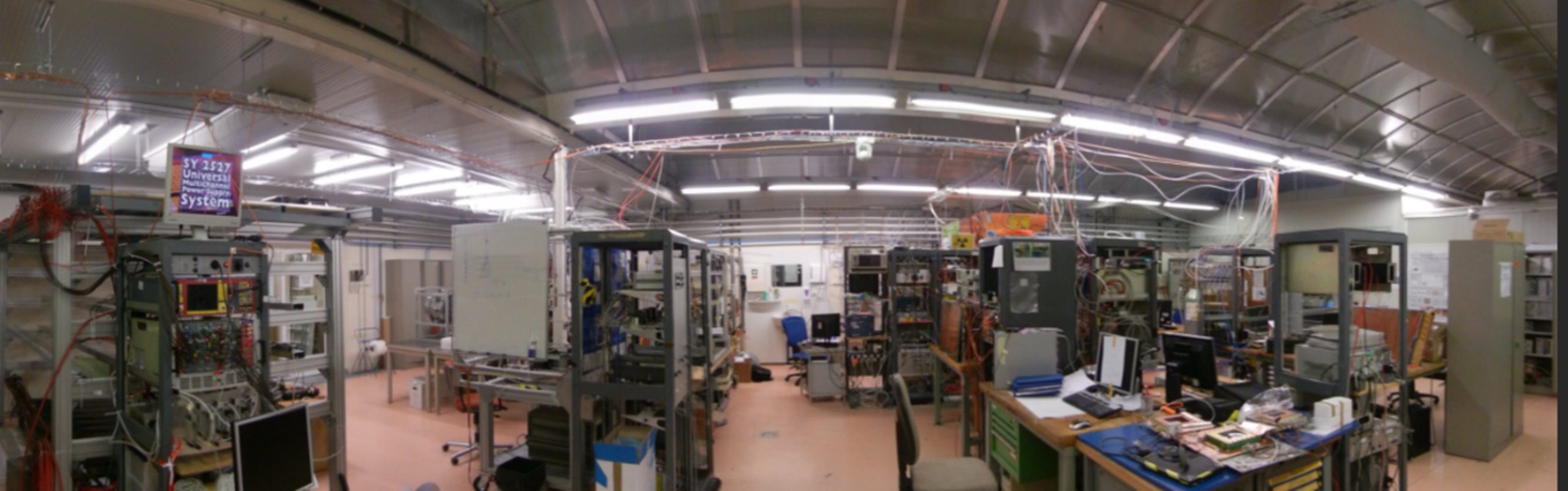}
\end{center}
\caption{\label{fig:GDDLaboratory}
\it{CERN EP-DT-DD Gaseous Detector Developments (GDD) laboratory.}}
\end{figure}

\begin{figure}[ht]
\begin{center}
\includegraphics[width=1\textwidth]{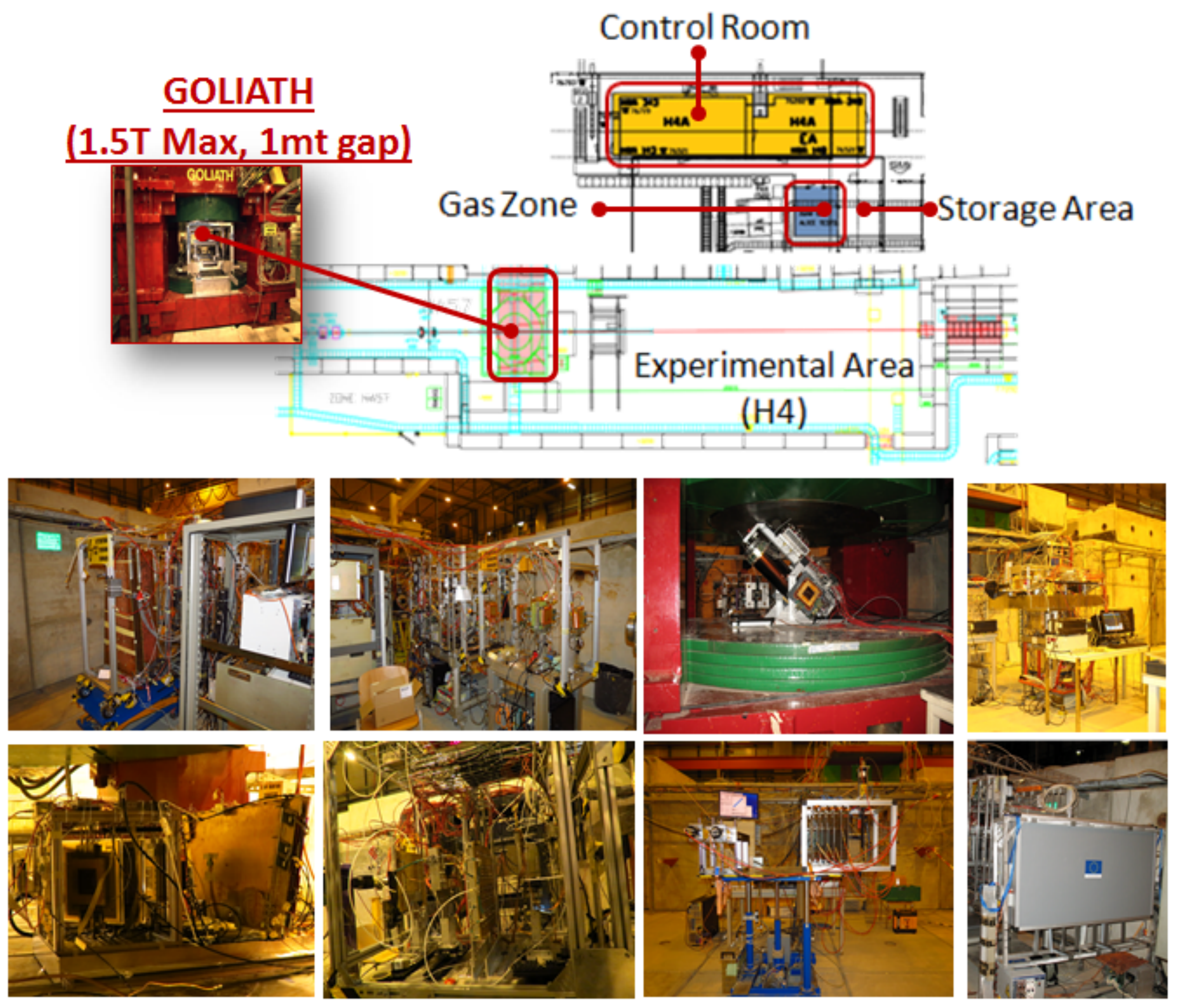}
\end{center}
\caption{\label{fig:TestBEAM}
\it{Top: Layout of the semi-permanent test beam installation for RD51 in the SPS Extraction Lines of the North Area . Bottom: Several setups equipped in the H4 line during a test beam campaign.}}
\end{figure}

Common spaces have been prioritized by the collaboration because of their important role in the process of sharing knowledge and expertise. The CERN EP-DT Gaseous Detector Development (GDD) laboratory has been enlarged and modernized and is accessible to the whole RD51 community (Fig.~\ref{fig:GDDLaboratory}). It is equipped with infrastructures, instrumentation and electronics needed for gaseous detector R\&D, such as radioactive sources, X-rays generators, gas-mixing units, standard and flammable gases, etc. Clean rooms are accessible for assembly, modification and inspection of the detectors.  It is the place where some common R\&D activities are carried out by different groups and institutions, making use of both, permanent and dedicated temporary installations. All three major LHC upgrades, incorporating MPGDs, started their R\&D in close contact with RD51, using dedicated setups at the GDD/RD51 laboratory. \\
    A semi-permanent common test-beam infrastructure has been installed at the H4 test beam area at CERN's SPS for the needs of the RD51 community (see Fig.~\ref{fig:TestBEAM}). The H4 area has the advantage that there is the "Goliath" magnet (around 1.5 T over a large area), allowing tests of MPGDs in a magnetic field. Collaboration members extensively use this facility for studying the performance of their new detectors. Three yearly periods, of two weeks each, have been always granted to RD51, with an average presence of four groups per period. It represents an efficient way of using the beam line. The presence of common test-beam infrastructure (e.g. high-precision beam telescopes, trigger detectors, HV control, DAQ, fiber optics lines, signal and Ethernet cables, stainless steel pipes, gas-distribution systems, etc.), eases installation and permits groups to rapidly start their experiments. Experience sharing and exchanges between groups is probably one of the most important and invaluable aspects of this common activity. The common GDD-RD51 laboratory provides important additional support, because it is located close to the test-beam area. Prototypes can be thoroughly scrutinized and tested before, during and after the beam campaign. The same argument is valid for the proximity of the Micro Pattern Technology Workshop. While the CERN common test facilities are the most valuable within the RD51 framework, each participating institute is willing to share facilities - thus creating a worldwide network of invaluable infrastructures. 
The extension of the collaboration will allow us to preserve the current framework and to enlarge possibly the existing support and network.

%% file: extension.tex
Numerous MPGD developments and applications summarized above, carried within the framework of RD51, have demonstrated that the collaborative R\&D and exchanges made within a large community of experts have led to a significant progress on all fronts over the past decade. Technological and scientific advances have led to improved detector performances - also when scaling up the new technologies to large-area detectors. This progress has been particularly crucial in HEP-experiments upgrades worldwide; examples were described above on the incorporation of GEM, MM, THGEM and their hybrid combinations in several large experiments. Also in the next years, 
a collaborative
R\&D phase and the right environment will have a strong impact on
project-oriented activities - similarly to the current scenario where three of the
major upgrades for the LHC experiments benefited from the RD51 framework.
The RD51 collaboration has been playing a crucial role in this 
progress - boosting exchanges and cooperation within the community on both scientific and technological matters. Common facilities and availability of experts and expertise have greatly contributed to the very noticeable progress. Frequent meetings, workshops and conferences permitted educating a young generation of detector experts - in laboratories situated all over the world. It is of a prime importance to preserve the expertise
that can be applied to other future detectors concepts. 
RD51 extension will permit further educational activities - paving the way towards a new, highly-motivated generation of young detector
scientists. Moreover, the collaboration will play a crucial role in attracting diversified resources.
CERN, collaborating institutions and projects contributions, industry,
EU projects, project synergies are a few examples of possibilities for the community.
\par
The better understanding of the physics processes, originating from experiment-validated model simulations, paved the way towards novel detector concepts. For example, higher granularities and rate capabilities, thinner detectors with embedded electronics, are expected to advance particle-flow based calorimetry (ILC-DHCAL) or high-precision tracking systems in high fluence environments. Fast and precise timing MPGDs are being developed for time tagging or filtering measurements, e.g. for high-luminosity colliders. Robust single-stage MPGD may pave the road to very large detection systems and civil imaging applications (e.g. radiography, homeland security); their industrial production would simplify the detector assembly procedures and possibly reduce the costs. High data-transfer bandwidth readout systems or optical readout can advance some potential applications. Hybrids detectors of combined technologies, e.g. gaseous and silicon, or gas and solid converter, can reach unprecedented performances. New concepts of charge and light sensing in noble-liquid detectors, may lead to progress in neutrino experiments and rare-event searches (e.g. Dark matter). Portable or sealed detectors would be obvious solutions to medical, cultural heritage, safety and security applications. Advanced materials and techniques can evolve into new sensitive or amplifying structures, extending the current detection sensitivities. This is just a partial list of fascinating R\&D lines that MPGDs will see in the years to come.
     
\subsection{R\&D Program on Advanced MPGD Concepts}
\textbf{Resistive materials and architectures}: originally introduced to improve space resolution, resistive materials and structures are another example where material science is very relevant to our field. Already used in several MPGD projects, resistive electrodes offer significant improvement in detector stability (discharge damping). It was shown that detectors with resistive anodes have led to several new structures incorporating single amplifying-stage MPGDs. Efforts will focus on the search for new resistive materials, with the right properties, radiation hardness and also the capability of operation under specific conditions (e.g. in dual-phase Ar and Xe detectors). New detector architectures will be investigated for efficiently damping eventual-discharge energy. DLC coating, originally introduced for the MSGC, is for instance at the base of the ongoing RWELL project: a single stage MPGD with single detector element. While different resistive-electrode architectures have been investigated, the proposed novelty is in design of more robust, easily-assembled and industrially-produced detector elements directly coming from the production process. The effects on large-area detector systems or for possible commercial use would be invaluable.
For example, in the context of precision tracking at high rates, the ongoing Small-Pad MM project is proposing DLC coatings for resistive MM, with high-granularity pads (few square mm) readout.  The proposed solution has been inspired by R\&D activities of the ongoing RD51 SCREAM project with a goal to construct the first MPGD-based sampling calorimeter prototype using resistive MM and RPWELL (THGEM-based) as active elements. Resistive structures like the embedded resistor and the RPWELL have the unique features that the resistive layer can be segmented into pads (no crosstalk due to surface diffusion) and the avalanche charge is evacuated to ground over short distances and therefore quickly (short RC time-constant). Fast charge clearance prevents charge buildup in the avalanche region and preserves the intrinsic proportional response of MPGDs at high rate and/or to high energy deposits. 

{\textbf{Fast and precise timing}}: Timing performances are often combined with other requirements, e.g. radiation hardness, spatial resolution and rate capabilities - imposing limits to technological choices. The PICOSEC project~\cite{TFG-picosec}, supported by RD51, has recently achieved, in its prototype version, time resolutions of $\sim$25psec for MIPs; the detector combines Cherenkov radiator, CsI photocathode and a MM sensor. The interest expressed by the community and the spectrum of potential applications makes this project an excellent candidate for further R\&D. It contains many aspects, common also to other detector concepts: among them, new converter and photosensitive materials, new fast-multiplier concepts, electronics, radiation hardness etc. This challenging project requires close interactions and synergies with other communities. Among them are Large Area Picosecond Photo-Detectors (LAPPD)~\cite{LAPPD} or FAST Fast Advanced Scintillator Timing ~\cite{FAST}. Novel fast electronics will get high priority in this project. The current RD51 activities on electronics are proving how much a collaboration environment can make the difference (resources, teamed-up needs, exchange of information and competencies). Radiation-hard fast scintillators and adequate photocathodes will require interactions with experts in the fields. Common interests of course would be with gaseous-based photon detectors for RICH counters and more generally with developments of Gaseous-Photo-Multipliers (GPM). The R\&D would necessitate the development of new strategies, new materials, new techniques and new architectures - expected to reduce, if not completely eliminate, the existing limitations. Entering such new fields of research will benefit from the existence of knowhow and expertise within our large collaboration.

\textbf{New materials and technologies}: material science is entering strongly in our field and can provide alternative MPGD materials, new structures, protection layers, etc. RD51 intends investigating MicroElectroMechanical Systems(MEMS)-like, nano-production techniques and 3D printing with conductive and insulating materials. The latter one could speed up particularly detector prototyping. The collaboration will create the proper environment for this type of developments, linking all members to the adequate infrastructures. Material studies for low out-gassing (contamination and aging), for longevity (radiation hardness), radio purity (rare events), efficiency (neutrons, photons), robust converters (single photons, UV), gas studies (eco-gases) will be a mandatory step to face. Previous experience (back to wires and MSGC) shows how in these kinds of studies a collaborative effort can offer good results in an efficient way. New knowledge has to be accumulated and the requested extension provides important support in this context. Low out-gassing materials would have impact on sealed detectors operation and that incorporating chemically unstable converters and photocathodes. Examples are: flame detection, radio protection, n-detection, medical applications, homeland security, space and cultural heritage (paintings, pyramids). The community's good background will be extended to cope with all the new possibilities.

\textbf{Hybrid detectors}: one characteristic property of 
current MPGDs is the hybridization
capability, namely the combination of different multiplication 
concepts, where RD51
members have investigated a variety of hybrid solutions. 
Such hybridizations are linked to specific technological 
competencies acquired along the development phases.
Hybrid detectors can offer improved avalanche-ion blocking:
COMPASS-RICH as an example where CsI-coated photocathode, THGEM
is followed by a MM multiplier. InGrid with the integration 
of MM on Timepix is another important example; it represents 
an ultimate performance for the ILC TPC: large MM signal, 
low noise, and extreme chip granularity. In the area of hybridization,
several successful ongoing R\&D projects are related to 
optical readout of gaseous
detectors; among them - optical avalanche imaging, 
beam monitoring fluorescence,
CT tomography, Compton camera and neutron detection, 
sensors for dual-phase TPCs. Current advances in digital cameras 
(CCD and CMOS in particular) are opening new possibilities of 
applications of low-cost, performing and ready to-use 
highly-pixelated readout systems. Even though high-resolution 
integrated imaging represents the most
common case, event-by-event and energy-resolved acquisitions 
are important in
some cases; an example is major interest in TPC readout 
(rare events, nuclear
physics). Here, hybrid design (charge and light detection) 
can be implemented,
combining the time evolution of the induced signal 
with high-resolution 2D projection.
In this application, future R\&D will 
clearly follow new cameras developments,
with improved sensitivity, faster acquisition and hopefully radiation-hard
imagers - thus paving the way to a large number of possible new applications.

%% file: assets_future.tex
\subsection{Support and Infrastructures}
\begin{figure}
\begin{center}
\includegraphics[width=0.8\textwidth]{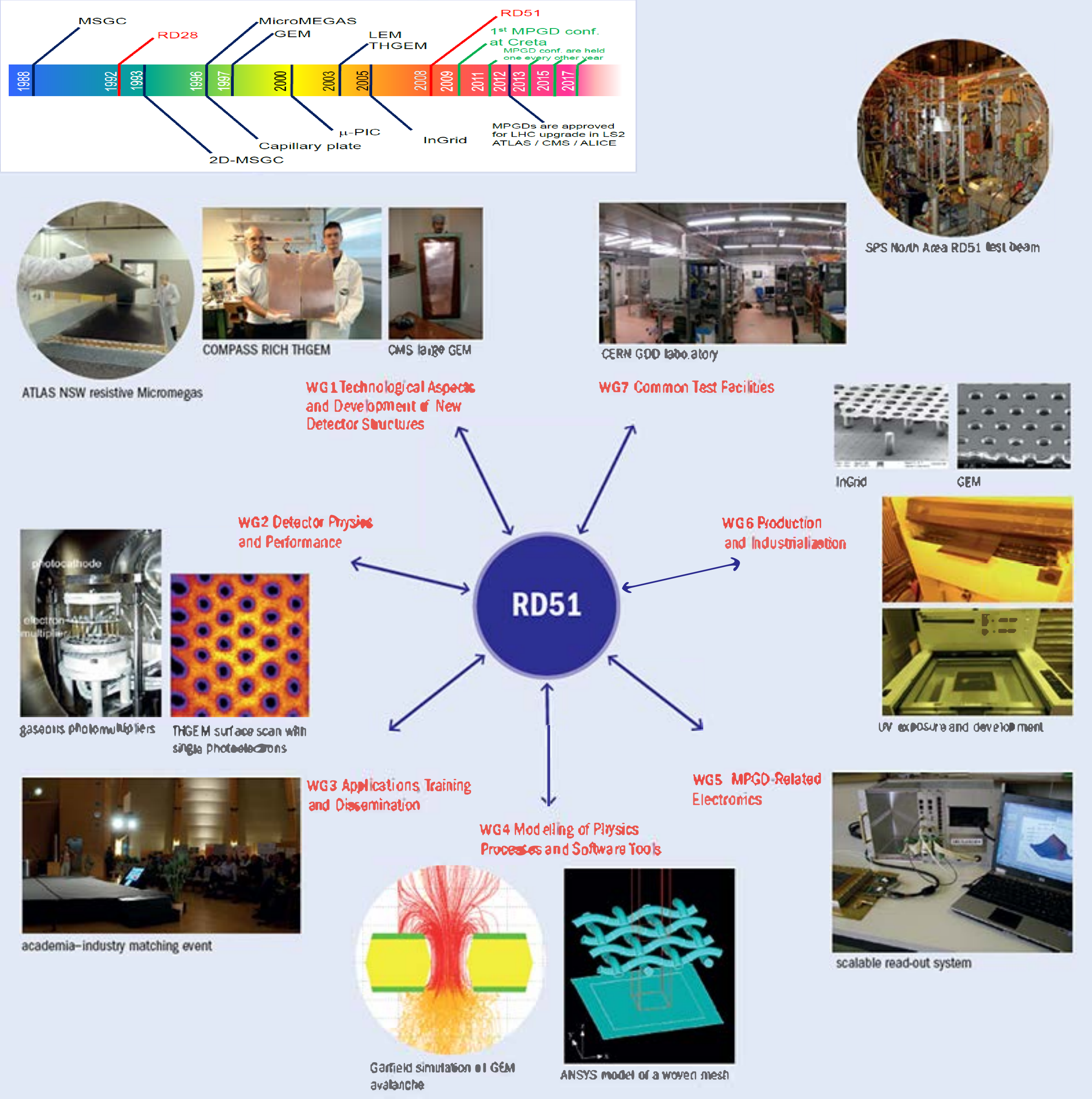}
\end{center}
\caption{\label{fig:RD51WorkingGroups}
The seven working groups of RD51, with illustrations of just a few examples of the different kinds of work involved. Top left: the 20-year pre-history of RD51.
Image credits: RD51 Collaboration.~\cite{rd51-cern-courier}}
\end{figure}

The collaboration is firmly interested on its extension for another five-year period to preserve and enrich the scenario presented in Sec.~\ref{assets_status}. The wide and international collaboration, built in the last ten years of activity,  is eager to continue contributing to the community. Support to generic R\&D,  wide expertise in MPGD technologies, unbiased exchange of knowledge and information, sharing of efforts are a few example of natural motivation towards future activities of the existing community.\\ 
The collaboration plans to maintain the current organization (Fig.~\ref{fig:organization}) and the existing structure based on the  working groups (Fig.~\ref{fig:RD51WorkingGroups}): 
\begin{figure}
\begin{center}
\includegraphics[width=0.8\textwidth]{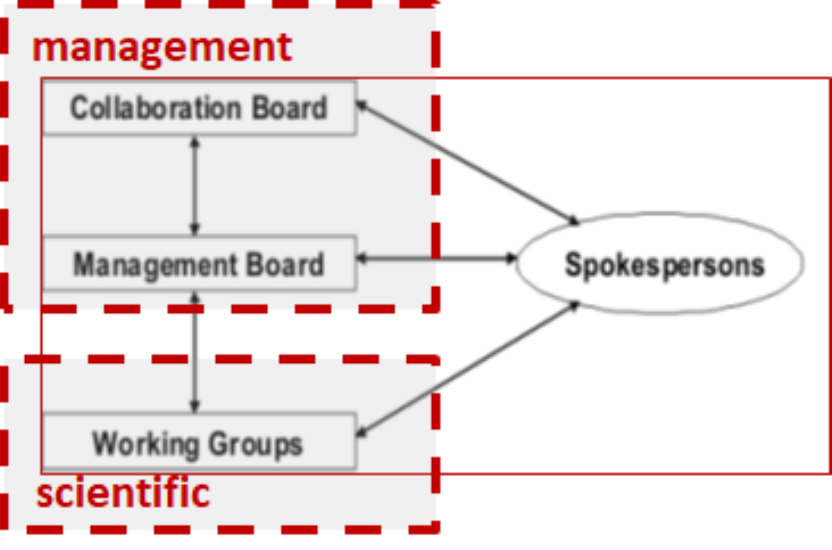}
\end{center}
\caption{\label{fig:organization}
The internal organization of the RD51 Collaboration is schematically illustrated.}
\end{figure}
\begin{itemize}
\item WG1 - Technological Aspects and Development of New Detector Structures
\item WG2 - Common Characterization and Physics Issues
\item WG3 - Applications
\item WG4 - Simulations and Software Tools
\item WG5 - MPGD Related Electronics
\item WG6 - Production 
\item WG7 - Common Test Facilities
\end{itemize}
New important areas of expertise will be covered, in particular in the field of novel and future technologies. The current status of support and infrastructures described in section ~\ref{assets_status} will be discussed in this section looking forward to the extension of the collaboration.\\

\textbf{{Detector physics, simulations and software tools}}: have been introduced in section~\ref{Simulation}. Support of Garfield++ package, which allows detailed simulation of small-scale structures, represents an important patrimony of RD51; it is an invaluable tool not only for MPGDs, but also for gaseous and silicon detectors, in general. The RD51 prolongation will allow extending present modeling efforts to new and complex processes useful for the community. It will permit extending the existing network of developers and users, forming a new generation of young experts and thus strengthening the portfolio of available tools. The collaboration also aims at supporting emerging developments to provide better simulation methods to describe the physics processes in MPGD detectors. For example, microscopic tracking algorithms in Garfield++ were devised and have shed light on the effects of surface and space charge in GEMs, as well as on the transparency of MM meshes. The collaboration will support ongoing efforts on interfacing between different modeling tools, started since a few years within the RD51 community. The aim is to address properly and efficiently most of the involved processes at the microscopic level: primary ionization, transport and avalanche, ions and ions clustering, charging up and space charge, signal induction and discharges. GEANT4~\cite{GEANT4}, COMSOL~\cite{COMSOL} and SPICE ~\cite{SPICE} are just a few examples. Note that these complete simulation platforms are of invaluable importance in the process of reaching the limit of a technology and detector performances.  Last, but not least, the extension will strongly support the presence of a core of experts available for developing, maintaining and transferring the existing knowledge. \\

\textbf{Electronics}: developments and detector-electronics integration advances have been described in section ~\ref{Electronics}. RD51 has been strengthening its expertise in the electronics domain for MPGDs since its origin. The extension will allow us to move forward toward novel exciting activities, to maintain and enhance existing support and to keep active the existing community.
   The major success of the RD51 collaboration - the SRS architecture - will continue to be used as a non-expensive and open platform for small and medium-sized systems. Many groups continue contributing to the SRS hardware, firmware and software maintenance and the system has already extended beyond RD51. It will be upgraded with new additional features to satisfy the needs of new challenging applications and projects. Within the Horizon2020 projects BrightnESS and AIDA2020, RD51 has decided to implement the VMM ASIC developed by BNL for the ATLAS NSW upgrade. The SRS-VMM will become the backbone of R\&D in MPGDs for the next decade and replace the successful, but discontinued, APV25-based readout, which drove the research within RD51 in the last eight years. The interface process, facilitated by the original SRS design and by the acquired expertise with the APV25, is ongoing. Input protection circuit, noise reduction, reading configuration and powering schemes are a few examples of the available know-how existing in RD51. Several groups have signed up to apply SRS-VMM in their experiments; e.g. for GEM-based readout at the NMX macromolecular diffractometer at the European Spallation Source.
    Another example originates from large-area, high granularity, pad readout detectors requiring proper readout solutions, where embedded electronics is an attractive possibility. This solution is feasible only if the know-how of properly integrating electronics and detector exists in the community. Challenging new developments will be also supported, such as the readout electronics for precise timing (tens of picoseconds) detection with MPGD. Front-end ASICs and SRS are clearly part of the new trends and there will be other developments and applications. RD51 will support these activities, improve testing capabilities, and contribute to reinvent mainstream technologies under a new paradigm of integration of electronics and detectors, as well as integration of functionality.    
The extension of the collaboration will allow our community to consolidate expertise, sustain the close interaction between electronics and detector experts, boost new developments and guarantee a general approach to satisfy a wider broader set of requests on detector-electronics integration, for a general use.\\

\textbf{Workshops, novel technologies and material}:
The extension of the collaboration for the next term will allow strengthening ongoing collaborative partnerships between RD51 and  Workshops; multidisciplinary collaboration is the stew in which creativity and innovation thrive. 
Synergies with the upgraded \textit{MPT workshop} at CERN will continue to be further enhanced. One has to highlight the role played by the community to advance MPGD production, in view of not only the LHC upgrades, but also for the benefit of other physics programs performed worldwide and potential commercial applications that may arise. Novel detector structures, stimulated by the needs and inspired by the community, are produced on the basis of acquired technological and scientific knowhow. The outcome of performance studies, quality control, and rapid feedback to production-process tuning, are just a few examples of contributions. The latter provides proper evaluation of the maturity of the technology. The possibility of rapid prototyping to access new solutions and production processes, in conjunction with the possibility to go for small, if not single production, facilitates the transfer of new detector technologies towards applications. An extension of the RD51 collaboration will secure an open, mutually reviewed, widespread and unbiased exchange of information and play a crucial role for application of new techniques, processes and materials.
{\textit{ Thin-film  Deposition Laboratories }}: a large interest in new techniques and materials is growing in the community. The extension of RD51 for another five-years term would give an important support for preserving and possibly enlarging this expertise ; it will motivate other new challenging activities (e.g. improvement of photocathode lifetime in gaseous detectors). New and more robust photocathodes, secondary electron emitter, thin resistive or protection layers - are just a few examples of future activities of interest in our community.  In this context, existing infrastructures and expertise are expected to grow in the field of material science and thin-film deposition,  will play an important role for the new developments.
{\textit{Material Science}}: a prolongation of  RD51 collaboration will allow directing efforts towards understanding and studying new detector-related materials and concepts. With sometimes unique and very peculiar properties, they might be good candidates for new and maybe exotic future developments in detection techniques. A wide range of possibilities can arise from these studies. For example, new types of radiation converters could enhance detection efficiencies; new amplifying structures could be eventually optimized (e.g. GEM made out of crystallized glass~\cite{GlassGEM} or glass piggyback Micromegas~\cite{PiggyBack}). Nanotechnologies are daily entering in various fields of science; one of the novel approaches is the study of the charge transfer properties through graphene deposited on electrodes in gaseous detectors~\cite{Graphene}. Some research activities in the collaboration have been already directed into this new R\&D.   The collaboration would like to direct efforts in bringing and facilitating the flow of knowledge and expertise from other communities to ours.\\

\textbf{Common space and common test facilities}: the development of robust and efficient MPGDs entails the understanding of their performance and implies a significant investment for laboratory measurements and detector test-beam activities to study prototypes and qualify final designs, including integrated system tests. Therefore, maintenance of the GDD (CERN PH/DT) lab at CERN and test-beam facilities at H4 SPS extraction line plays a key role among the objectives of RD51. The extension of the collaboration will allow us to preserve the current framework and possibly enlarge the existing support and scientific network.

%% file: requests.tex
Since the initial period of RD51, namely since 2009, the support offered by CERN facilities has substantially contributed to the collaboration activities and achievements - without major impact on CERN resources. A similar support is requested for the next term; it consists of:
\begin{itemize}
\item access to the GDD lab space, infrastructure and maintenance support;
\item office space and administrative support;
\item maintenance of the semi-permanent setup at the SPS H4 test beam line and, correspondingly, access to the beams over several time periods for a total of six weeks per year;
\item continuation of the collaborative access to the:
\begin{itemize}
\item the EP-DT-EF MPT (Micro Pattern Technology) Workshop;
\item the EP-DT-EF TFG (Thin Film and Glass) Laboratory;
\end{itemize}
\item access to other CERN technical facilities, in particular:
\begin{itemize}
\item EP-DT-DD Bond Laboratory
\item TS-DEM-WS Electronics Assembly Workshop
\item EN-MME-MM Materials, Metrology \& NDT
\item TE-VSC Surface treatment, coating and chemical analysis
\item the central computing resources for MPGD simulations.
\end{itemize}
\end{itemize}
The development of the next generation of MPGDs and, more in general, of ionizing particle detectors can largely profit of novel emerging technologies as those related to nanomaterials, 
MicroElectroMechanical Systems (MEMS), sputtering, 
novel photoconverters, 3-D printing options, 
as also tested within our planning for future R\&D discussed in Sec.~\ref{extension}. 
It is expected that, both at CERN and within the R\&D community, 
these new needs will result in the formation of synergies 
and networks between scientists devoted to detector R\&D and groups 
and institutions mastering the novel technologies. In selected sectors, this process can also result in novel infrastructural resources at CERN. 
Therefore, RD51, while stimulating the community of the detector 
developers towards the exploration of the emerging technologies, 
require accessing the dedicated networks and possible future 
infrastructures. Furthermore, in a context of constructive 
synergies, RD51 is willing to offer support to the actions 
towards the dissemination of emerging technologies for detector 
developments and assembly.